\documentclass[aps,prb,groupedaddress,amsmath]{revtex4}

\usepackage{amssymb}
\usepackage{amsmath}
\usepackage{amscd}
\usepackage{graphicx}
\usepackage{amsbsy}
\usepackage{color}

\begin{document}
\title{Phase diagram and structural properties of a simple model for one-patch particles}
\author{Achille Giacometti}
\affiliation{Dipartimento di Chimica Fisica, Universit\`a di Venezia,
Calle Larga S. Marta DD2137, I-30123 Venezia, Italy}
\email{achille@unive.it}
\author{Fred Lado}
\affiliation{Department of Physics, North Carolina State University, Raleigh, North Carolina 
27695-8202}
\email{fred_lado@ncsu.edu}
\author{Julio Largo}
\affiliation{Departamento de F\'{\i}sica Aplicada, Universidad de Cantabria, Avenida de los Castros
s/n, Santander 39005, Spain}
\author{Giorgio Pastore}
\affiliation{Dipartimento di Fisica Teorica, Universit\`a di Trieste  
and CNR-INFM Democritos,
Strada Costiera 11, 34151 Trieste, Italy}
\email{pastore@ts.infn.it}
\author{Francesco Sciortino}
\affiliation{{Dipartimento di Fisica and
  INFM-CNR-SOFT, Universit\`a di Roma {\em La Sapienza}, Piazzale A. Moro  2, 00185 Roma, Italy}}

\date{\today}
\begin{abstract}
We study the thermodynamic and structural properties of a simple, one-patch fluid model using the reference hypernetted-chain (RHNC) integral equation and specialized Monte Carlo simulations. In this model, the interacting particles are hard spheres, each of which carries a single identical, arbitrarily-oriented, 
attractive circular patch on its surface; two spheres attract via a simple square-well potential only if the two patches on the spheres face each other within a specific angular range dictated by the size of the patch. For a ratio of attractive to repulsive surface of $0.8$, we construct the RHNC fluid-fluid separation curve and compare with that obtained by Gibbs ensemble and grand canonical Monte Carlo simulations. We find that RHNC provides a quick and highly reliable estimate for the position of the fluid-fluid critical line. In addition, it gives a detailed (though approximate) description of all structural properties and their dependence on patch size.
\end{abstract}

\maketitle
\section{Introduction}
\label{sec:intro}
The role that anisotropic interactions play in the structure of a fluid system and on its equilibrium phase diagram is the subject of intense 
current investigation; a renewed interest is reinvigorating previous studies that attempt to understand and model the structure and dynamics of 
associating liquids. \cite{Wertheim84,Nezbeda05,Carlevaro03,Varakin97,Jirsak07}
This renewed interest is stimulated by two complementary recent developments. First, profound advances in chemical particle synthesis are significantly augmenting the naturally available choices, offering the possibility of controlling surface patterns and functionalities in colloidal particles. \cite{Glotzer07} Command of the assembly properties of these new patchy colloidal particles will give fine control over the three-dimensional organization of materials as well as the combination of different materials over several length scales, promising to yield a spectrum of crystal polymorphs and assembled structures that is unprecedented in colloid science. And second, growing evidence that to understand protein interactions in aqueous solution and to describe quantitatively their phase diagrams could require accounting for the presence of anisotropic protein-protein interactions. Aeolotopic \cite{Lomakin99,McManus07} interactions [from the Greek aiolos (variable) and topos (place)] are now beginning to be considered fundamental \cite{Pellicane08,Liu07} for  grasping the nature and mechanism of self-aggregation processes in protein systems, particularly in fibril and crystal formation.

In the case of spherically symmetric potentials, a rather clear understanding has been reached of the topology of the phase diagram, of the 
structural properties of the liquid phase, and of the possible crystal structures. Much less understood are highly anisotropic models, where the 
local structure of such fluids is still rather poorly predicted. This is
particularly true in those cases in which the anisotropy of the interaction or its patchiness produces first-nearest neighbor shells with low 
coordination number. Most of the evidence for the structure of these systems arises from numerical simulations, \cite{Horsch05,Bianchi06,Bianchi07,Teixeira00,Fantoni07,Foffi07} which is only rarely accompanied by quantitative theoretical predictions. In this respect, primitive models of patchy particles \cite{Nezbeda05}  --- by condensing the interparticle repulsion into a hard-core potential while attractive interactions are modeled as simple square-well potentials --- offer the possibility of establishing general trends and acting as test cases for checking novel theoretical approaches or extensions to the anisotropic case of familiar approximations whose validity has already been assessed for spherical potentials.

One interesting primitive model is the Kern-Frenkel potential, \cite{Kern03} based upon an idea introduced by Chapman {\it et al.} \cite{Chapman88} 
in the framework of molecular fluids. In the simplest realization of this model, which we discuss in this article, the surface of the particle is 
partitioned into two parts, one with square-well and one  with hard-sphere character. The relative ratio between attractive and repulsive  surfaces 
is measured as coverage $\chi$. Two particles attract if they are within a specific distance of each other and if their attractive surfaces are 
properly aligned with each other. 

Here we present an integral equation study of the phase diagram and of the structural properties of the one-patch Kern-Frenkel potential \cite{Kern03} for a specific value of the coverage ($\chi=0.8$) based on the  reference hypernetted-chain (RHNC) approximation introduced in 
Ref. \onlinecite{Lado73} for spherical potentials and later extended to molecular fluids. \cite{Lado82a,Lado82b}  This approach provides, besides thermodynamic  information, an extremely detailed description of the structure of the liquid via a pair distribution function which now includes both radial and angular correlations, i.e., with a level of detail which can hardly be reached in experiments or in simulations. We compare the theoretical estimates of the gas-liquid coexistence and of appropriately-averaged radial distribution functions with ``exact'' numerical simulations of the model to assess the ability of the RHNC approach to describe such highly anisotropic interactions. 

As the size of the patch decreases, lower temperatures are required to follow the fluid-fluid coexistence line and integral equation convergence becomes progressively harder to achieve. Still, in the fluid region, it is possible to provide a detailed description of patch size dependence for the pair distribution function. We find a rich physical behavior which cannot be captured by 
rotationally averaged quantities, such as those computed by numerical simulations, nor by any effective temperature rescaling.

Attractive interactions need not be restricted to short-range square-well potentials. In a very recent paper, G\"ogelein \textit{et al.} \cite{Goegelein08} used a hard-sphere Yukawa potential, within a similar patchy model, to incorporate possible longer-range interactions due to the addition of salt, but they used second-order thermodynamic perturbation theory to derive the phase diagram, in contrast with the integral equation approach followed here.

The paper is organized as follows. In Section \ref{sec:ie}, we introduce the one-patch Kern-Frenkel model used in this work, while in Section \ref{sec:oz} we describe the main steps necessary for the solution of the RHNC integral equation of the model, stressing analogies and differences with similar approaches already exploited in the framework of molecular fluids; a few technical details are collected in the Appendices. Section \ref{sec:therm} deals with the computation of thermodynamics and coexistence lines. Finally, Section \ref{sec:numerical} contains the RHNC results along with their comparison with Monte Carlo simulations and Section \ref{sec:conclusions} points out some conclusions and draws some perspectives.  

\section{The model}
\label{sec:ie}

\begin{figure}[htbp]
\vskip0.5cm
\includegraphics[width=8cm]{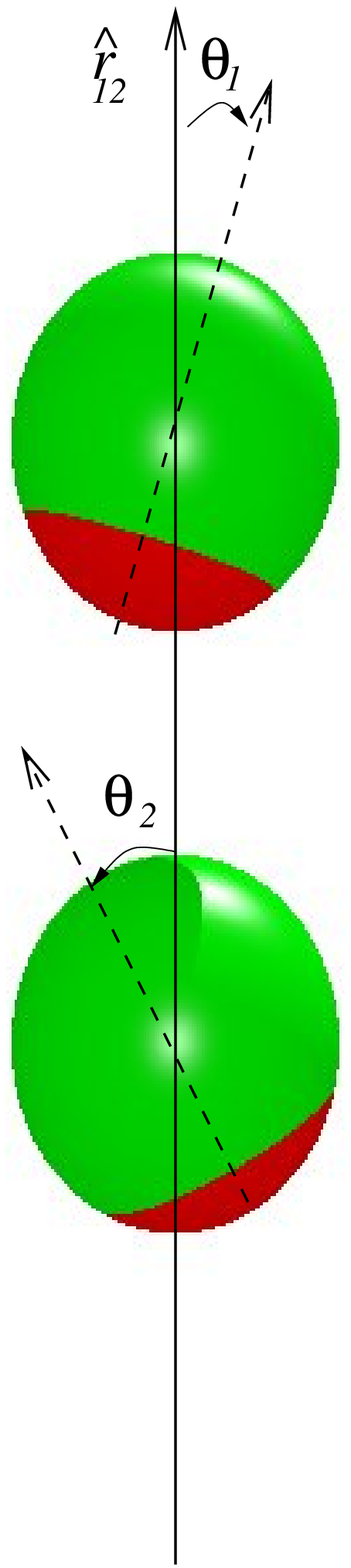} \\
\vskip0.5cm
\includegraphics[width=8cm,angle=0]{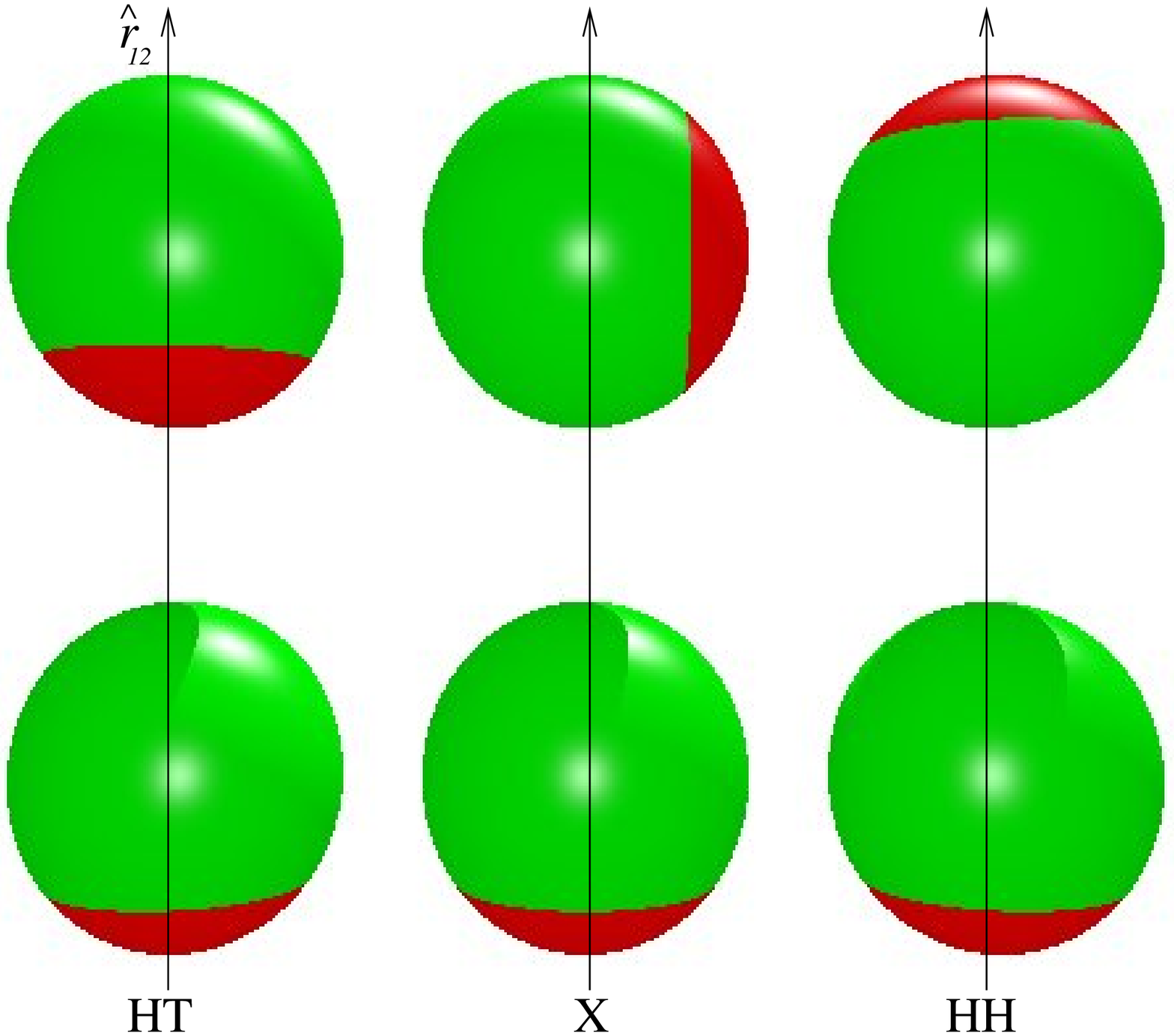}
\caption{The one-patch Kern-Frenkel potential, where $\hat{\mathbf{r}}_{12}$ is the direction joining the two centers (top panel). 
Directions of the patches 
are specified by unit vectors $\hat{\mathbf{n}}_1$ and $\hat{\mathbf{n}}_2$. Of particular relevance are the relative orientations of the two patches as defined 
by $\theta_{12}=\theta_2-\theta_1$: $\theta_{12}=0$ (HT), $\theta_{12}=\pi/2$ (X), and $\theta_{12}=\pi$ (HH) (bottom panel). Note that a larger, lighter spherical 
surface corresponds to the attractive part, meaning here a coverage $\chi>0.5$.
\label{fig:kf_pot}}
\end{figure}

We study the Kern-Frenkel patchy hard sphere model, \cite{Kern03} patterned after an idea introduced by Chapman {\it et al.} \cite{Chapman88} in the framework of molecular fluids: Two spherical particles attract via a short-range square-well potential only if two patches on the respective surfaces are properly aligned (see Fig. \ref{fig:kf_pot}).

In the case of a {\em single} patch per particle, the pair potential reads \cite{Kern03} 
\begin{eqnarray}
\Phi\left(12\right) &=& \phi \left(r_{12}\right) \Psi\left(\hat{\mathbf{n}}_1,
\hat{\mathbf{n}}_2,\hat{\mathbf{r}}_{12} \right),
\label{ie:eq1}
\end{eqnarray}
where
\begin{equation}
\phi\left(r\right)= \left\{ 
\begin{array}{ccc}
\infty,    &  &   0<r< \sigma    \\ 
- \epsilon, &  &   \sigma<r< \lambda \sigma   \\ 
0,          &  &   \lambda \sigma < r \ %
\end{array}%
\right.  \label{ie:eq2}
\end{equation}
and
\begin{equation}
\Psi\left(\hat{\mathbf{n}}_1,
\hat{\mathbf{n}}_2,\hat{\mathbf{r}}_{12}\right)= \left\{ 
\begin{array}{ccccc}
1,    & \text{if}  &   \hat{\mathbf{n}}_1 \cdot \hat{\mathbf{r}}_{12} \ge \cos \theta_0 & \text{and} &  
-\hat{\mathbf{n}}_2 \cdot \hat{\mathbf{r}}_{12} \ge \cos \theta_0,   \\ 
0,    &  & &\text{otherwise}, & \
\end{array}%
\right.  \label{ie:eq3}
\end{equation}
$\theta_0$ being the angular semi-amplitude of the patch. Here $\hat{\mathbf{n}}_1(\omega_1)$ and $\hat{\mathbf{n}}_2(\omega_2)$ are unit vectors 
giving the directions of the center of the patch in spheres 1 and 2, respectively, with $\omega_1=(\theta_1,\varphi_1)$ and 
$\omega_2=(\theta_2,\varphi_2)$ their corresponding spherical angles in an arbitrarily oriented coordinate frame. Similarly, 
$\hat{\mathbf{r}}_{12}(\Omega)$ is the unit vector of the separation $r_{12}$ between the centers of mass of the two spheres and is defined by the 
spherical angle $\Omega$. As usual, we have denoted with $\sigma$ and $\lambda \sigma$ the hard core diameter and the width of the well.

Patchy models are the minimal version of models used to describe systems with highly directional attractions and have recently been applied to a variety of anisotropic cases, ranging 
from network and associated fluids, \cite{Kolafa87,Varakin97,Demichele1,Demichele2} colloids, \cite{Foffi07,Fantoni07} proteins, \cite{Kern03,Liu07,Goegelein08} and gels. \cite{Bianchi07,Russo09}

\section{Solution of the Ornstein-Zernike equation}
\label{sec:oz}

Our integral equation approach is based on the reference hypernetted-chain (RHNC) approximation introduced in Ref. \onlinecite{Lado73} for spherical potentials and later extended to molecular fluids. \cite{Lado82a,Lado82b} In Appendix \ref{app:rhnc}, we briefly summarize the necessary key expressions. 
Within this scheme, the closure equation takes on the assumed-known bridge function $B_0(12)$ of a particular reference system to replace the actual 
unknown bridge function $B(12)$ appearing in the exact closure (see Eq. (\ref{app:eq2})). 
Parameters of $B_0(12)$ are treated as adjustable, a viewpoint initiated by Rosenfeld and Ashcroft, \cite{Rosenfeld79} who postulate that the form of the bridge function $B(r)$ for simple fluids constitutes the same univeral family of curves, irrespective of the pair potential, and that the known hard sphere version --- with appropriate choice of the hard sphere diameter $\sigma_0$ --- can serve as a reference model for all of them. They demonstrate this hypothesis by explicit calculation, \cite{Rosenfeld79} selecting the optimum $\sigma_0$ in each case so as to yield close agreement with published simulation results. \cite{comment} It is in fact possible to optimize the reference system parameters autonomously via a variational free energy principle \cite{Lado82} that enhances internal thermodynamic consistency, in that Hiroike's condition, \cite{Hiroike57}
\begin{equation}
\rho \frac{\partial}{\partial\rho} \left( \frac{\beta U}{N} \right) = \beta \frac{\partial}{\partial\beta} \left( \frac{\beta P}{\rho} \right),
\end{equation}
is recovered. This follows because, as in the original HNC equation, the internal energy $U$ and pressure $P$ are again obtained as thermodynamic derivatives of the same free energy functional.

Anisotropic integral equations are far more demanding than their spherical counterparts from a computational point of view, but the basic procedure, involving expansions of the angular dependence in spherical harmonics, has been tested for numerous physical systems \cite{Lado95,Lombardero99,Gray84} with potentials having a continuous dependence on the angular orientations. The unusual {\em discontinuous} nature of the angular dependence appearing in Eq. (\ref{ie:eq3}), however, would seem to present a real challenge to such a spherical harmonics expansion, given especially that theoretically infinite expansions become necessarily finite in numerical work. In fact, there is no need to expand this potential at all; it may be used directly as-is within a specified angular grid. 

We briefly sketch now the main steps of the integral equation procedure. The iterative solution of the Ornstein-Zernike (OZ) equation plus an 
approximate closure equation typically requires a series of back and forth Fourier transformations between functions in direct and momentum space. Besides the bridge function $B(12)$, this involves the direct correlation function 
$c(12)$, the pair distribution and correlation functions $g(12)$ and $h(12)=g(12)-1$, respectively, and the auxiliary function $\gamma(12)=h(12)-c(12)$ (see Appendix \ref{app:rhnc}).
 
Angular dependence introduces additional back and forth Clebsch-Gordan transformations between the coefficients of angular expansions in ``axial'' 
frames, with $\hat{\mathbf{z}}= \hat{\mathbf{r}}$ in direct space or $\hat{\mathbf{z}} = \hat{\mathbf{k}}$ in momentum space, and more general ``space'' frames with arbitrarily-oriented axes. (Note that here 
and in much of the following we have put for brevity $\hat{\mathbf{r}}\equiv \hat{\mathbf{r}}_{12}$.) This is now a well-established procedure whose general scheme is summarized in Table I. We note that, because of  the cylindrical symmetry of the angular dependence in Eq. (\ref{ie:eq3}), the simpler version of the procedure for linear molecules \cite{Lado82a} rather than the full anisotropic version \cite{Lado95} can be exploited here.

We start an iteration with the current values of the axial coefficients $\gamma_{l_{1} l_{2} m}(r)$ in $\mathbf{r}$ space (see Appendix \ref{app:rhnc}) and use the expansion in spherical harmonics, Eq. (\ref{app:eq3}) with $X=\gamma$, to construct $\gamma(r,\omega_1,\omega_2)$. 
The bridge function $B(r,\omega_1,\omega_2)$ is similarly constructed in RHNC approximation, while $\Phi(r,\omega_1,\omega_2)$ is given. Then using the closure equation (\ref{app:eq2}) to obtain $c(r,\omega_1,\omega_2)$ and the expansion inverse Eq. (\ref{app:eq4}) (with $X=c$) to compute its axial coefficients $c_{l_{1} l_{2} m}(r)$, the algorithm proceeds by a Clebsch-Gordan transformation to obtain the corresponding space coefficients $c(r;l_{1} l_{2} l )$ from Eq. (\ref{app:eq6}). This is followed by a Hankel transform, Eq. 
(\ref{app:eq8}), to momentum space and a backward Clebsch-Gordan transformation, Eq. (\ref{app:eq7}), to return to an axial frame in $\mathbf{k}$ 
space, yielding $\tilde{c}_{l_{1}l_{2} m}(k)$. The OZ equation (\ref{app:eq5}) in momentum space can now be invoked to get $\tilde{\gamma}_{l_{1} l_{2} m}(k)$. This is then followed by a forward Clebsch-Gordan transformation, Eq. (\ref{app:eq6}), and an inverse Hankel transform, Eq. (\ref{app:eq9}), to provide $\gamma(r;l_{1} l_{2} l)$. A final backward Clebsch-Gordan transformation, Eq. (\ref{app:eq7}), brings a new estimate of the original coefficients $\gamma_{l_{1} l_{2} m}(r)$. 

This cycle is repeated until self-consistency between input and output $\gamma_{l_{1} l_{2} m}(r)$ is achieved. 

\section{Thermodynamics}
\label{sec:therm}
\subsection{Internal energy and virial pressure}
\label{subsec:energy_pressure}

The internal energy $U$ of the system can be computed starting from the standard expression \cite{Gray84}
\begin{eqnarray}\label{energy_pressure:eq1}
U &=&  \left \langle \sum_{i=1}^{N} \sum_{j>i}^N \Phi\left(ij \right) \right \rangle
=\frac{1}{2} \rho N \int d \mathbf{r}_{12} \left \langle g\left(12\right) 
\Phi\left(12\right) \right \rangle_{\omega_{1} \omega_{2}}.
\end{eqnarray}
Here $\rho$ is the number density and $\langle \ldots \rangle_{\omega}=(1/4\pi)\int d \omega \ldots$  indicates an average over the 
spherical angle $\omega=(\theta,\varphi)$. This expression can be written in a simpler form by exploiting the factorization in Eq. (\ref{ie:eq1}):
\begin{eqnarray}
\label{energy_pressure:eq2}
\frac{\beta U}{N} &=& \frac{1}{2} \rho \int d \mathbf{r} ~\beta \phi\left(r\right)
G\left(r\right) = -2\pi \rho \beta \epsilon \int_{\sigma}^{\lambda \sigma} dr ~ r^2
G\left(r\right),
\end{eqnarray}
where $\beta=1/k_B T$ is the inverse Kelvin temperature (with $k_B$ Boltzmann's constant) and
\begin{eqnarray}
\label{energy_pressure:eq3}
G\left(r\right) &=& \left \langle g(r,\omega_1,\omega_2 \right) \Psi\left(\omega_1, \omega_2 \right) \rangle_{\omega_1 \omega_2} 
\equiv \frac{1}{\left(4 \pi\right)^2} \int d \omega_1 d \omega_2~ 
g\left(r,\omega_1,\omega_2\right) \Psi\left(\omega_1, \omega_2 \right). 
\end{eqnarray}
The angular integrals in Eq. (\ref{energy_pressure:eq3}) are evaluated using Gaussian quadratures \cite{Lado82a} while the trapezoidal rule is used for Eq. (\ref{energy_pressure:eq2}). (The average number of neighbors in the well is found as $\bar{z}=-2U/N\epsilon$.)

The virial pressure $P$ can also be obtained from a standard expression, \cite{Gray84}
\begin{eqnarray}
\label{energy_pressure:eq7}
P &=& \rho k_B T - \frac{1}{3V} \left \langle \sum_{i=1}^N \sum_{j>i}^N
r_{ij} \frac{\partial}{\partial r_{ij}} \Phi\left(ij \right)
\right \rangle =\rho k_B T - \frac{1}{6} \rho^2 \int d \mathbf{r}_{12}
\left \langle g\left(12\right) r_{12} \frac{\partial}{\partial r_{12}} \Phi\left( 12 \right)
\right \rangle_{\omega_{1} \omega_{2}}.
\end{eqnarray}
Using familiar manipulations, the definition of the continuous cavity function 
$y(12)=g(12) e^{\beta \Phi(12)}$, and the result
\begin{eqnarray}
\label{energy_pressure:eq8}
\frac{\partial}{\partial r_{12}} \left[ e^{-\beta \phi \left(r_{12}\right)
\Psi\left(\omega_1,\omega_2\right)} \right] &=& 
e^{\beta \epsilon \Psi\left(\omega_1,\omega_2\right)} \delta\left(r_{12}-\sigma \right)
- \left[ e^{\beta \epsilon \Psi\left(\omega_1,\omega_2\right)} -1 \right]
\delta\left(r_{12}-\lambda \sigma\right),
\end{eqnarray}
Eq. (\ref{energy_pressure:eq7}) can be cast in the form
\begin{eqnarray}
\label{energy_pressure:eq9}
\frac{\beta P}{\rho} &=& 1+ \frac{2}{3} \pi \rho \sigma^3 \left \{ 
\left \langle y\left(\sigma,\omega_1,\omega_2 \right) 
e^{\beta \epsilon \Psi\left(\omega_1,\omega_2\right)}\right \rangle_{\omega_1 \omega_2}
-  \lambda^3 \left \langle y\left(\lambda \sigma,\omega_1,\omega_2 \right) \left[
e^{\beta \epsilon \Psi\left(\omega_1,\omega_2\right)}-1 \right] \right \rangle_{\omega_1
\omega_2} \right \},
\end{eqnarray}
which again can be reduced to Gaussian quadratures.

Finally, the isothermal compressibility $\kappa_T= \rho^{-1} (\partial \rho/\partial P)_T$ can
be obtained from \cite{Gray84}

\begin{eqnarray}
\label{energy_pressure:eq10}
\rho k_B T \kappa_T &=& 1+\rho \int ~d\mathbf{r} \left[g_{000}\left(r\right) -1 \right].
\end{eqnarray}

\subsection{RHNC free energy}
\label{subsec:free_energy}

It has been shown \cite{Lado82b} that in RHNC approximation, just as with its parent HNC equation, \cite{Morita60} the excess free energy 
per molecule can be 
calculated in a self-contained fashion as
\begin{eqnarray}
\label{free_energy:eq1}
\frac{\beta F_{\rm ex}}{N} &=& 
\frac{\beta F_{1}}{N} +\frac{\beta F_{2}}{N}+\frac{\beta F_{3}}{N},
\end{eqnarray}
where
\begin{eqnarray}
\label{free_energy:eq21}
\frac{\beta F_1}{N} &=& -\frac{1}{2} \rho \int d \mathbf{r}_{12} 
\left \langle \frac{1}{2} h^2\left(12\right)+h\left(12\right)-g\left(12\right) \ln \left
[ g\left(12\right) e^{\beta \Phi\left(12\right)} \right] \right \rangle_{\omega_{1} \omega_{2}}, \\
\label{free_energy:eq22}
\frac{\beta F_2}{N}&=& - \frac{1}{2 \rho} \int \frac{d\mathbf{k}}{\left(2\pi\right)^3}
\sum_{m} \left \{ \ln \mathrm{Det} \left[ \mathbf{I} + \left(-1\right)^m \rho
\tilde{\mathbf{h}}_m \left(k\right) \right] - \left(-1\right)^m
\rho \mathrm{Tr} \left[\tilde{\mathbf{h}}_m \left(k\right) \right] \right \}, \\
\label{free_energy:eq23}
\frac{\beta F_3}{N}&=& \frac{\beta F_3^0}{N}-\frac{1}{2} \rho \int d \mathbf{r}_{12} \left \langle \left[ g\left(12\right)-g_0\left(12\right) 
\right] 
B_0\left(12\right) \right 
\rangle_{\omega_{1} \omega_{2}}.
\end{eqnarray}
In Eq. (\ref{free_energy:eq22}), $\tilde{\mathbf{h}}_m(k)$ is a Hermitian matrix with elements $\tilde{h}_{l_1 l_2 m}(k)$, $l_1,l_2 \ge m$, and $\mathbf{I}$ is the unit matrix. The last equation, for $F_3$, directly expresses the RHNC approximation. Here $F_3^0$ is the reference system contribution, computed from the known free energy $F^0_{\rm ex}$ of the reference system as $F_3^0=F^0_{\rm ex}-F_1^0-F_2^0$, with $F_1^0$ and $F_2^0$ calculated as above but with reference system quantities. 
 
The bridge function $B_0(12)$ appearing in (\ref{free_energy:eq23}) is taken from a reference system and is the key approximation in the RHNC scheme, since it replaces the unknown bridge function $B(12)$ in the general closure equation (\ref{app:eq2}). At present only the hard sphere (HS) model is sufficiently well known to play the role needed here, and so we put $B_{0}(12)=B_{\rm HS}(r_{12};\sigma_0)$. The optimum hard sphere diameter $\sigma_0$ is selected 
according to a variational free energy minimization that yields the condition 
\cite{Lado82}
\begin{eqnarray}
\rho \int d \mathbf{r} \left [ g_{000}\left(r\right) - g_{\rm HS}
\left(r;\sigma_0\right) \right]  \frac{\partial B_{\rm HS} \left(r;\sigma_0\right)}{\partial \sigma_0}&=&0.
\label{free_energy:eq5}
\end{eqnarray}
Note that $\sigma_0=0$ corresponds to an HNC closure while $\sigma_0=1$ corresponds to the true hard-sphere diameter of the potential. In actual numerical computations, the value of $\sigma_0$ clearly will depend upon the considered state points and coverage. This dependence of $\sigma_0$ in our results is further discussed in Section \ref{subsec:ie}.
\subsection{Chemical potential}
\label{subsec:chem_pot}

The chemical potential $\mu$ is conveniently obtained within the RHNC scheme from the relation
\begin{eqnarray}
\label{chem_pot:eq1}
\beta \mu = \frac{\beta F}{N} + \frac{\beta P}{\rho},
\end{eqnarray}
which does not introduce any additional approximation into the calculation. The ideal quantities are
$\beta F_{\rm id}/N = \ln (\rho \Lambda^3)-1$, $\beta P_{\rm id}/\rho=1$, and $\beta \mu_{\rm id} = \ln (\rho \Lambda^3)$,
where $\Lambda$ is the de Broglie wavelength.

We stress the fact that, within the RHNC scheme, the chemical potential directly follows from the free energy and pressure, themselves computed directly from the integral equation solution for the pair functions, unlike other closures where additional approximations are required. This is a particularly important feature when numerical results stemming from different closures are compared. 

\section{Numerical results}
\label{sec:numerical}
\subsection{Monte Carlo simulations}
\label{subsec:mc}

We have studied the equilibrium properties of the one-patch model with several standard Monte Carlo methods. We perform Gibbs ensemble simulations \cite{Frenkel02} 
to evaluate coexistence in the region where the gas-liquid free-energy barrier is sufficiently high to avoid crossing between the two phases. Here we use a system of 1200 particles which partition themselves into two boxes whose total volume is $4300\sigma^3$, corresponding to an average density of $\rho \sigma^3=0.27$. At the lowest $T$ this corresponds to roughly $1050$ particles in the liquid box and 150 particles in the gas box (of side $\approx 13\sigma$). On average, the code attempts one volume change every five particle-swap moves and 500 displacement moves. Each displacement move is composed of a simultaneous random translation of the particle center (uniformly distributed between $\pm 0.05 \sigma$) and a rotation (with an angle uniformly distributed between $\pm 0.1$ radians) around a random axis. To calculate the location of the gas-liquid critical point, we perform grand canonical Monte Carlo (GCMC) simulations, \cite{Frenkel02} complemented with histogram reweighting techniques to match the distribution of the order parameter $\rho - s e$ with the known functional dependence expected at the Ising universality class critical point. \cite{Wilding97} Here $e$ is the potential energy density, $\rho$ the number density, and $s$ the mixing field parameter. We have studied systems of size $7\sigma$ up to $10\sigma$ to estimate the size dependence of the critical point. We did not perform a finite-size study since the changes in critical parameters with the size are not significant. In GCMC we attempt, on average, one insertion/deletion step every 500 displacement steps. We have also performed a set of GCMC simulations for different choices of $T$ and $\mu$ to evaluate  the corresponding density. Finally, we performed standard MC simulations of a system of 1000 particles in the NVT ensemble to 
estimate structural properties for comparison with integral equations results. 

\subsection{Integral equations}
\label{subsec:ie}

We solve the RHNC equations numerically on $r$ and $k$ grids of $N_r=2048$ points, 
with intervals $\Delta r = 0.01 \sigma$ and $\Delta k = \pi/(N_r\Delta r)$, using a standard Picard iteration method. \cite{Hansen86} The square-well width is 
set at $\lambda=1.5$ because of results available for the isotropic square well. \cite{Giacometti09} For the hard-sphere bridge function, we use the 
Verlet-Weis-Henderson-Grundke parametrization. \cite{Verlet72,Henderson75} Reduced temperature $T^*=k_B T/\epsilon$ and density $\rho^*=\rho \sigma^3$ are used throughout. 

The angles $\theta_j$, $j=1,2,\ldots,n$, that appear in the Gauss-Legendre quadratures over polar angle $\theta$ are determined by the $n$ zeroes of the Legendre polynomial $P_n(\cos\theta)$ of order $n$; it is not possible in general to ensure that one of these will coincide with the patch 
parameter $\theta_0$. To obtain a ``best fit,'' we have chosen $n$ in the following {\em ad hoc} fashion. The total coverage $\chi$ (fraction of surface covered by the attractive patch) can be obtained from Eq. (\ref{ie:eq3}) as (cf. Ref. \onlinecite{Kern03})
\begin{eqnarray}
\label{num:eq1}
\chi &=& \left\langle \Psi\left(\hat{\mathbf{n}}_1,\hat{\mathbf{n}}_2,\hat{\mathbf{r}}_{12} 
\right) \right \rangle_{\omega_1 \omega_2}^{1/2} = \sin^2\left(\frac{\theta_0}{2}\right).
\end{eqnarray}
By also {\em numerically} evaluating the angular average in this expression using Gauss-Legendre quadrature of order $n$ and varying $n$ (typically from 30 to 40), we find that number of Gaussian points $n$ that minimizes the error in the numerical computation of $\chi$, the exact value being known. Then for a fixed value of  $\chi$, all Gaussian quadratures in a calculation are evaluated with the same number $n$ of points. For the case $\chi=0.8$ studied in most detail in this article, we found $n=31$ to be the optimal value, yielding a $0.25\%$ error in Eq. (\ref{num:eq1}). For lower $\chi$ values, the value $n=30$ was used instead. 

Expansions in spherical harmonics are terminated after $l_1=l_2=m=4$, resulting in 35 distinct coefficients in a pair function expansion; we have explicitly checked that this suffices for all cases considered here. Hankel transforms have been obtained using a step-up/step-down procedure as detailed in Ref. \onlinecite{Lado82a}.
In all calculations, the covergence criterion was a root-mean-square difference between successive iterates of less than $10^{-5}$.
At $80\%$ coverage, typical values for the variational parameter $\sigma_0$ range from approximatively $0.77$ for reduced densities of order
$10^{-2}$ to $1.02$ in the liquid branch (reduced density $0.75$). These latter values decrease for lower coverages to reach the HS limit $1.0$ when $\chi \to 0$, as expected.

\subsection{Phase coexistence}
\label{subsec:phase}
\begin{figure}[htbp]
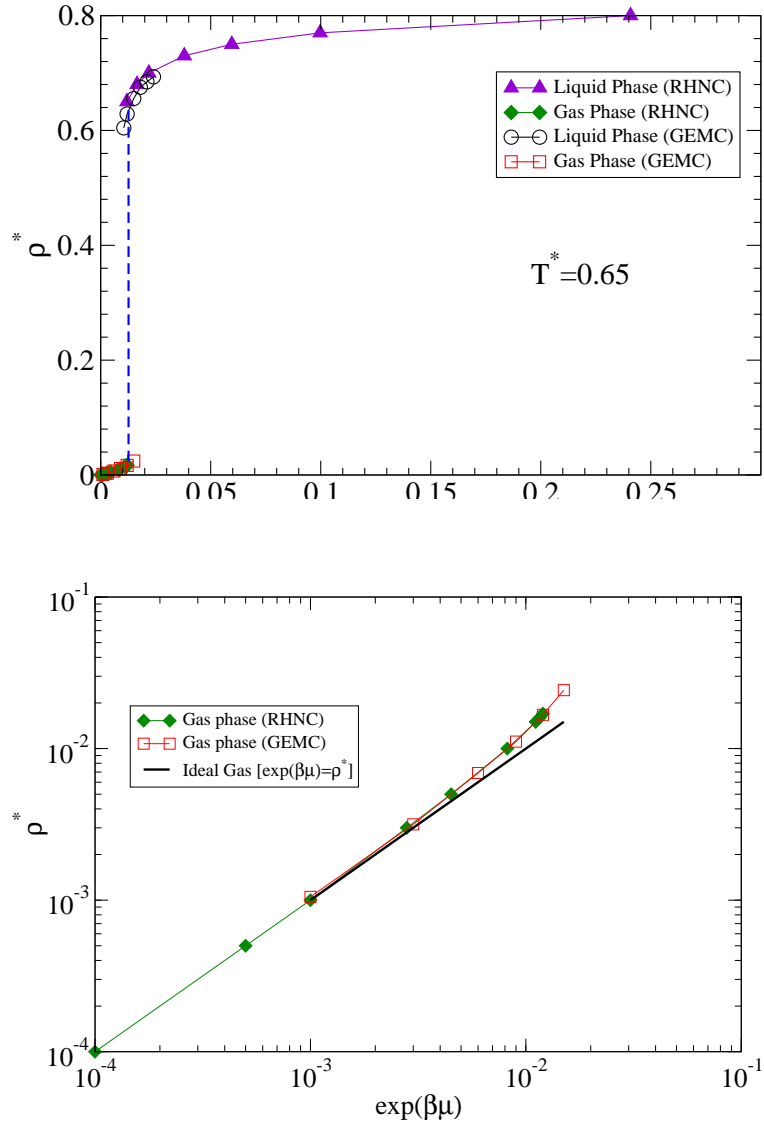

\includegraphics[width=10cm,angle=0]{Fig2a.eps} \\
\vskip0.5cm
\includegraphics[width=10cm,angle=0]{Fig2b.eps} 
\vskip0.5cm
\caption{Plot of reduced density $\rho^{*}$ versus $e^{\beta \mu}$ for $\chi=80\%$ coverage (top panel). 
Results are for both gas and liquid branches as computed 
from GCMC and RHNC. The bottom panel focuses only on the gas branch to enlarge the small-density region and display consistency with the correct ideal gas limit.
Table \ref{tab:expbetamu} provides the numerical values for both GEMC and RHNC calculations.
\label{fig:fig1}}
\end{figure}

For the computation of the phase diagram we have here considered a representative value of coverage $\chi=0.8$
Figure \ref{fig:fig1} depicts a comparison of the results for the reduced density $\rho^{*}$ as a function of the activity $e^{\beta \mu}$. 
Here $\mu$ is the total 
chemical potential, so that $\rho^{*}=e^{\beta \mu}$ is the ideal gas limit, where we have replaced the density-independent de Broglie wavelength $\Lambda$ with the hard sphere $\sigma$ for convenience. A comparison with GEMC results shows that RHNC reproduces the correct behavior both qualitatively and quantitatively  
for both the low (gas) and high (liquid) densities branches. This is a rather stringent test on the performance of RHNC since the chemical potential 
is usually a very sensitive quantity to the approximation used. 
For the liquid branch, the numerical values stemming from the two procedures are also reported in Table \ref{tab:expbetamu}.

Armed with this result we turn next to the computation of the phase diagram. There exist very few studies of the fluid-fluid coexistence line using integral equations even in the framework of spherical, fully isotropic, potentials. One possible scheme to identify the densities of the two coexisting phases was outlined by Schlijper \textit{et al}. \cite{Schlijper93} for a Lennard-Jones fluid. A similar method was used in Ref. \onlinecite{Giacometti09} for a square-well fluid. A more elaborate procedure has been devised in Refs. \cite{Charpentier05} and \cite{ElMendoub08}. Our analysis closely follows the spherical counterpart given in Ref. \onlinecite{Giacometti09} in order to make a close connection with the fully attractive case.
 
For a fixed temperature below the critical one, the gas and liquid branches are scanned along the density axis until the numerical program is unable to produce a converged solution. This might be due either to the disappearance of a solution of the integral equation itself, which is a well-known shortcoming \cite{Belloni93} of the hypernetted-chain closure and its descendants, or to the approach  of spinodal instability. Coexistence density values $\rho_g$ and $\rho_l$ for the chosen temperature are then obtained by the numerical solution of the system
\begin{eqnarray}
\label{phase:eq1}
P(\rho_l) &=& P(\rho_g), \\
\mu(\rho_l) &=& \mu(\rho_g).
\end{eqnarray}

\begin{figure}[htbp]
\vskip0.5cm
\includegraphics[width=10cm,angle=0]{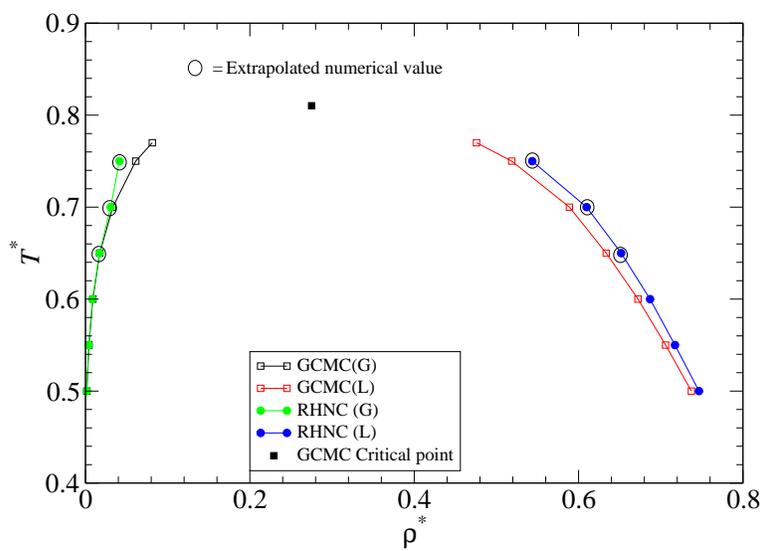} 
\caption{Phase diagram  for $\chi=0.8$ ($80\%$ coverage). Circled points indicate that an extrapolation procedure has been used in the RHNC data.
Table \ref{tab:coextable} provides the RHNC numerical values.
\label{fig:fig2}}
\end{figure}

It might (and it does) happen that approaching the spinodal line, it is not possible to obtain a solution of the integral equation before Eq. (\ref{phase:eq1}) can be satisfied. In such cases, a suitable small extrapolation compensates for this limitation of the approximate closure; details of this extrapolation process can be found in Ref. \onlinecite{Giacometti09}. Figure \ref{fig:fig2} depicts the phase diagram so obtained compared with GEMC results, where extrapolated points have been highlighted. We note that RHNC results follow remarkably well GEMC data along both branches.
For RHNC, the numerical values of the (reduced) densities and temperature along with the common values of chemical potentials
and pressure, as given by Eq.(\ref{phase:eq1}), are listed in Table \ref{tab:coextable}.
The small systematic deviation from the GEMC results in the high density branch is roughly of the same order of magnitude as that found in the SW case, \cite{Giacometti09} thus strongly suggesting that angular dependence has not worsened the results. This discrepancy would obviously be reduced by a better description of the anisotropic bridge function, about which essentially nothing is known. Minor deviations in the low density branch, on the other hand, are mostly due to numerical convergence problems, especially for temperatures close to the critical region, as indicated by the extrapolation marks.

As the coverage $\chi$ decreases, we find that the critical temperatures and densities shift towards progressively smaller values.
This happens also for higher number of patches and in general it cannot be accounted for by a simple rescaling of the temperature. \cite{Foffi07} 
We remark, however, that the one-patch case is very peculiar in this respect for two reasons. First, because it is the only case (along with its two-patch counterpart)
clearly interpolating between the full isotropic HS ($\chi=0$) and SW ($\chi=1$) limits. For 3 or more patches on each sphere, a well-defined coverage
limit (smaller than $100\%$) is found. Second, because of new peculiar phenomena arising for low coverages, notably for $\chi \le 0.5$.
This and other points will the the subject of a future publication.

\subsection{Pair correlation functions and structure factor for $\chi=0.8$ coverage}
\label{sec:pair_single}

Given the results obtained for the phase diagram, it is interesting to consider structural information from an analysis of correlation functions and structure factors. Unlike with their isotropic counterparts, information about pair distribution functions $g(12)$ with angular dependence is obtained differently in integral equation calculations and Monte Carlo simulations. Integral equations provide a complete, although approximate, description of the pair distribution function $g(12)$, which in the ``canonical'' axial frame is a function of four variables: $r_{12}$, $\theta_1$, $\theta_2$, and $\phi_{12}=\phi_2-\phi_1$. This level of detail is a prohibitive computational task for simulations, which instead produce a more accessible {\em average} of some sort. In the present simulations, we calculate an average over all possible orientations of the vector joining the centers of mass of the two spheres,  
$\hat{\mathbf{r}}_{12}(\Omega)$, while holding fixed the orientations of the individual vectors $\hat{\mathbf{n}}_{1}(\omega_{1})$ and $\hat{\mathbf{n}}_{2}(\omega_{2})$, as well as the magnitude $r_{12}$. This produces a function of just two variables: $r_{12}$ and $\theta_{12}=\theta_2-\theta_1$. 

\begin{figure}[htbp]
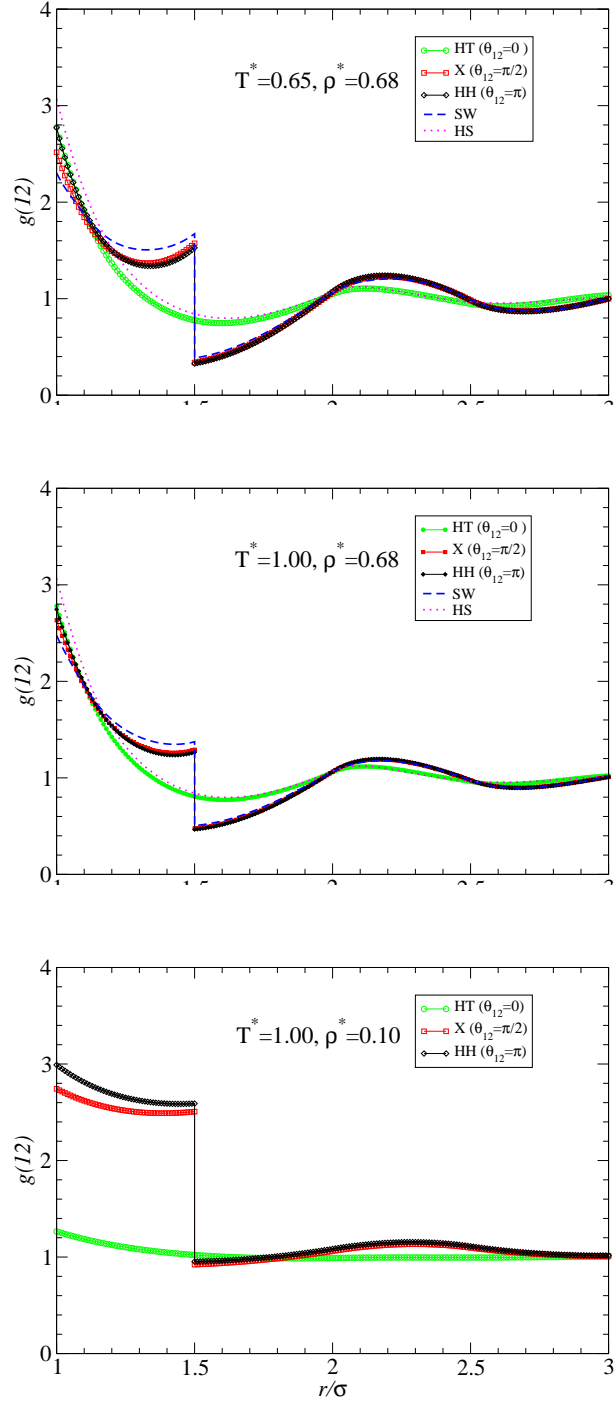

\includegraphics[width=8cm,angle=0]{Fig4a.eps}\\
\vskip0.5cm
\includegraphics[width=8cm,angle=0]{Fig4b.eps}\\
\vskip0.5cm
\includegraphics[width=8cm,angle=0]{Fig4c.eps} 
\caption{Plots of $g(12)$ as a function of $r/\sigma$ for $\rho^{*}=0.68$ and temperatures $T^{*}=0.65$ (top panel) and $T^{*}=1.0$ (center panel). 
Different lines  refer to three specific patch orientations having angles $\theta_{12}=0$ (HT), $\theta_{12}=\pi/2$ (X), and $\theta_{12}=\pi$ (HH), 
all at $\chi=0.8$ coverage. 
For comparison, in the bottom panel the corresponding low-density case $\rho^{*}=0.1$, $T^{*}=1.0$ is reported. 
SW and HS indicate the isotropic square-well and hard-sphere limits, respectively.
Note that all panels have been drawn on the same scale. 
\label{fig:fig3}}
\end{figure}

Detailed results for specific orientations from RHNC, {\em without} averaging, can be seen in Fig. \ref{fig:fig3}, where we plot $g(12)=g(r,\omega_1,\omega_2)$ as a function of $r/\sigma$ for two state points lying in the liquid-like region ($\rho^{*}=0.68$) with two different temperatures: $T^{*}=1.0$ above the critical point and $T^{*}=0.65$ in the subcritical region. The three different curves in each plot refer to three different values of the product $\hat{\mathbf{n}}_1 \cdot \hat{\mathbf{n}}_2 $; these are $\hat{\mathbf{n}}_1 \cdot \hat{\mathbf{n}}_2=1$ for a head-to-tail (HT) configuration, 0 for a crossed (X) configuration, and -1 for a head-to-head (HH) configuration.
 
Coverage has again been set at $80\%$ and the width of the well is as always $\lambda=1.5$. The jump occurring at $r/\sigma = 1.5$ in the HH and X configurations closely resembles the jump occurring in the isotropic counterpart, \cite{Giacometti09} indicating a clear bonding 
due to the attractive part of the potential. This does not occur for the HT configuration, where the patches are on opposite sides (see Fig.\ref{fig:fig1}). This feature is common to all coverages smaller than $100\%$ (i.e., a square-well potential) 
and it is related to the particular angular dependence of the Kern-Frenkel model, as given in Eq.(\ref{ie:eq3}), coupled with
the fact that $g(12)$ is here computed in the axial frame where $\hat{\mathbf{n}}_1$ is along the direction $\hat{\mathbf{r}}_{12}$.
Then $\hat{\mathbf{n}}_1 \cdot \hat{\mathbf{r}}_{12}=1$ and $-\hat{\mathbf{n}}_2 \cdot \hat{\mathbf{r}}_{12}=-1$ in the HT configuration thus yielding
no coupling unless $\theta_0=\pi$ corresponding to the isotropic square-well potential. 
Results corresponding to the isotropic square-well potential ($100\%$ coverage) and hard-sphere potential ($0\%$ coverage) are also depicted for contrast. 

A few features of the above behavior are worth noting. First the hard core contact values ($r=\sigma^{+}$) in the HH and HT configurations are virtually identical at both temperatures $T^{*}=1.0$ and $T^{*}=0.65$, whereas the X configuration is significantly different, this difference being greater for the lower $T^{*}=0.65$ temperature.

As temperature decreases from $T^{*}=1.0$ to $T^{*}=0.65$, the SW contact value decreases, the X contact value also decreases while the corresponding HH and HT ones are essentially unchanged. This directionality ordering is reversed at the other contact point inside the well ($r=\lambda \sigma^{-}$) where the SW value
is larger than both the HH and X corresponding ones, and this difference increases with increasing temperature. The HT configuration has a much lower value as there is no discontinuity in this case, as remarked. Inspection of the HT contact values is sufficient to convince that the head-to-tail configurations display a significant temperature dependence in spite of their hard-sphere-like apparent behaviors.

The above results can be understood by contrasting these cases with attractive interactions with the corresponding HS ones (independent of temperature). Here the $r=\sigma^{+}$ contact value is larger than all other cases (including the SW one) where the other $r=\lambda \sigma^{-}$ extreme has a much lower value compared to SW, HH and HT configurations. This is because attractive interactions decrease the probability of sphere-touching occurrence while it increases the probability of an indirect bonding within the well region
$\sigma <r < \lambda \sigma$ mediated by the presence of intermediate particles. For lower coverage, the overall attractive force is reduced and this effect is less marked and, more importantly, it depends upon the directionality of the bond.

A corresponding low-density ($\rho^{*}=0.1$, $T^{*}=1.0$) state is also reported 
in Fig. \ref{fig:fig3} in order to account for a gas-like behavior. Not surprisingly, in this case both HH and X configurations behave similarly, with HH configuration always lying above the corresponding X one, since for such a low density the low probability of finding nearest-neighbor spheres is mostly independent of the orientation (the fully isotropic results is not reported in Fig.\ref{fig:fig3} as this state point lies inside the critical line; see Ref. \onlinecite{Giacometti09}). 

\begin{figure}[htbp]
\includegraphics[width=6cm,angle=0]{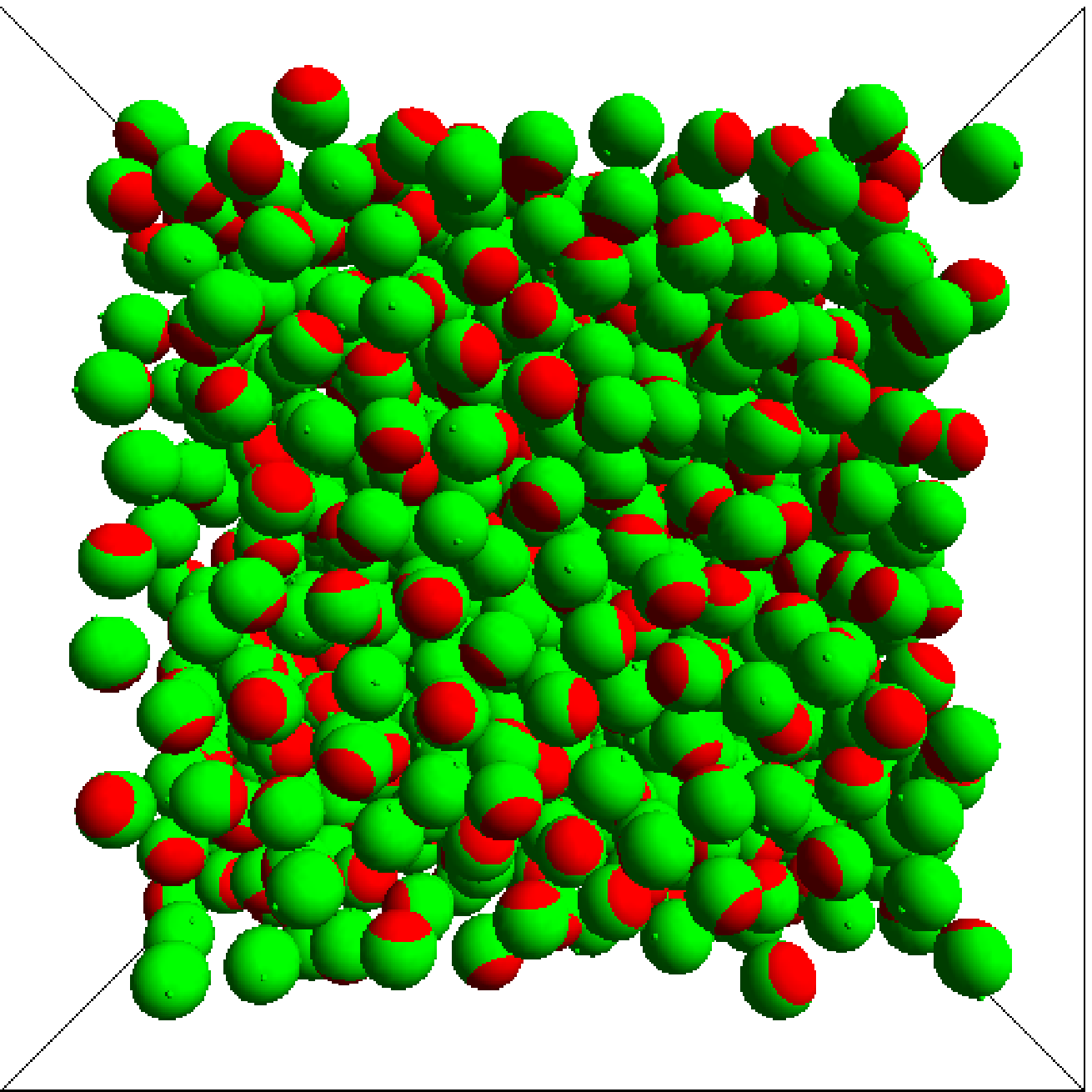}\\
\vskip0.5cm
\includegraphics[width=6cm,angle=0]{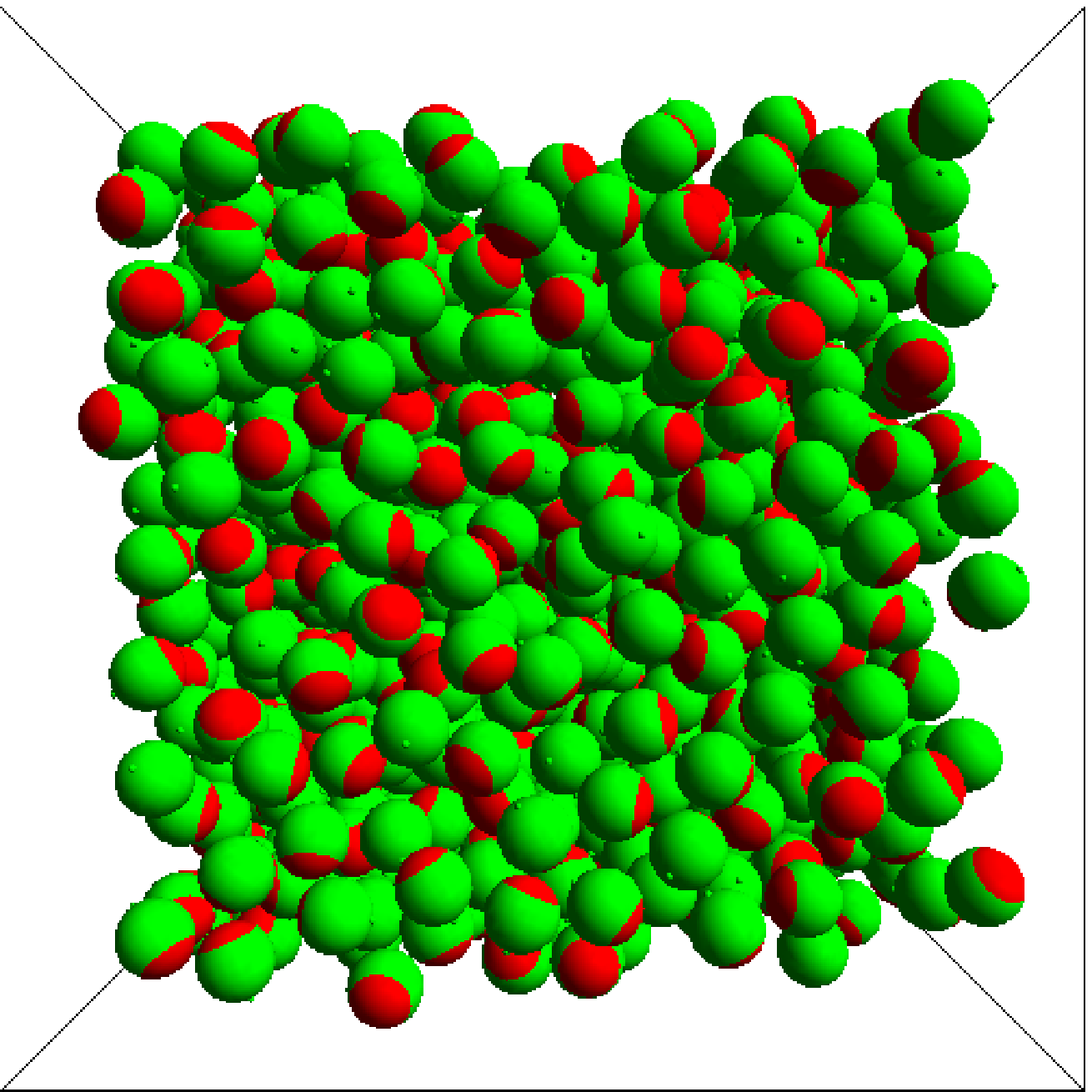} \\
\vskip0.5cm
\includegraphics[width=6cm,angle=0]{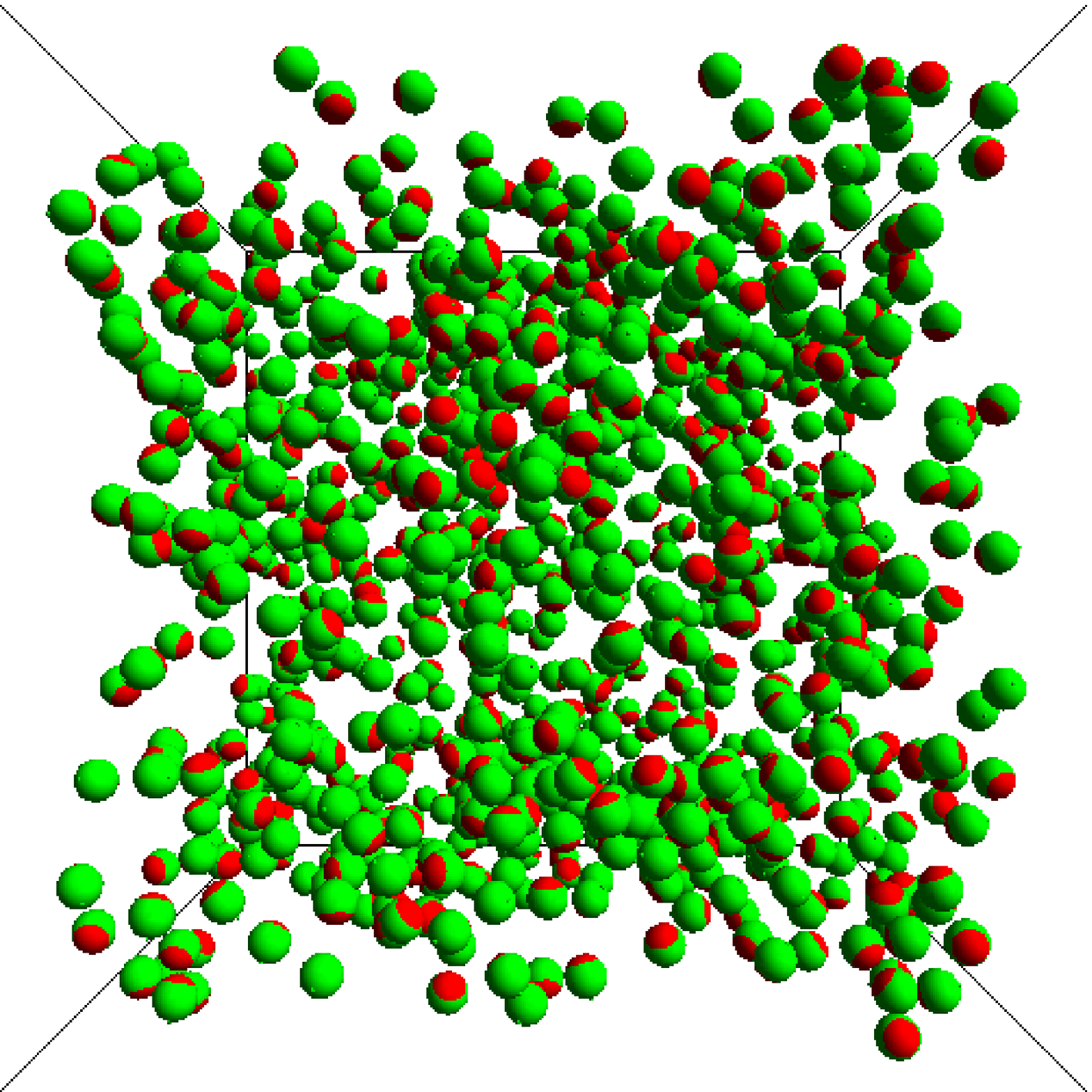} 
 \caption{(color online) Snapshots of two $80\%$ coverage configurations with $\rho^{*}=0.68$ and temperatures $T^{*}=0.65$ (top panel) 
and $T^{*}=1.0$ (center panel). 
The last panel depicts a corresponding low-density case with $\rho^{*}=0.1$, $T^{*}=1.0$. The three selected state points are the same values considered 
in previous figures.
\label{fig:fig4}}
\end{figure}

Snapshots of configurations for the above state points are displayed in Fig. \ref{fig:fig4}. The liquid-like structure is clearly visible in the first two configurations with $\rho^{*}=0.68$, without any significant morphological difference between the supercritical $T^{*}=1.0$ and the 
subcritical $T^{*}=0.65$ thermodynamic states. The RHNC provides also a prediction for the average number of particles in the well $\bar{z}$ 
(that is, with center-to-center distance within the range $[\sigma, \lambda \sigma ]$) by integrating the center-to-center radial distribution function
$\langle g(12)\rangle_{\omega_1,\omega_2,\Omega}$
within the well. In the above cases we find approximatively $7.5$ and $8.3$ for $T^{*}=1.0$ and $T^{*}=0.65$, respectively. 

Conversely, in the vapor-like state $\rho^{*}=0.1$, $T^{*}=1.0$, a much lower average number of particles, $1.6$, is found in the well, as expected. This figures have been obtained upon integration of the $g(12)$ over the radial range $\sigma <r <\lambda r$ averaged over all possible relative orientations.

\begin{figure}[htbp]
\includegraphics[width=6cm,angle=0]{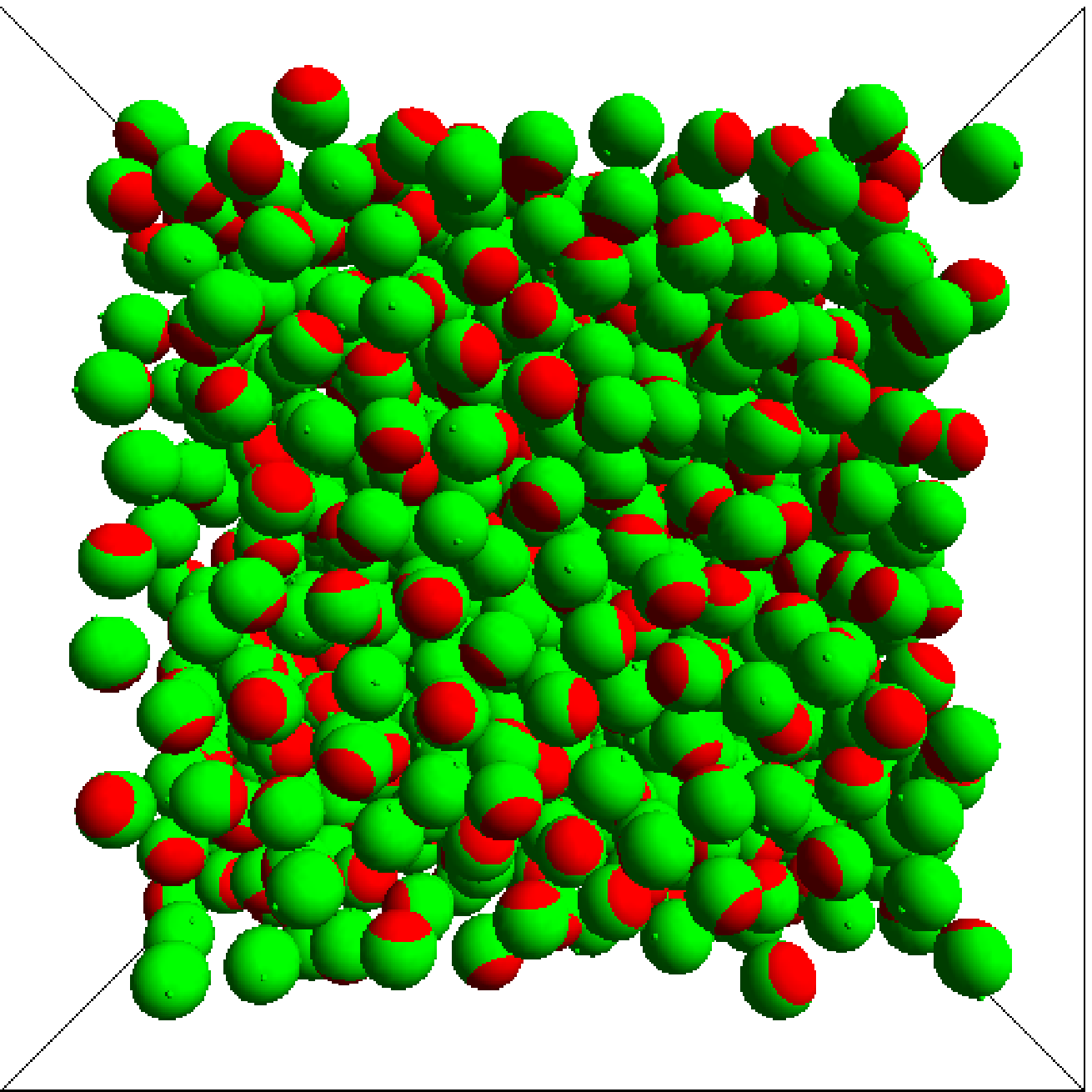}\\
\vskip0.5cm
\includegraphics[width=6cm,angle=0]{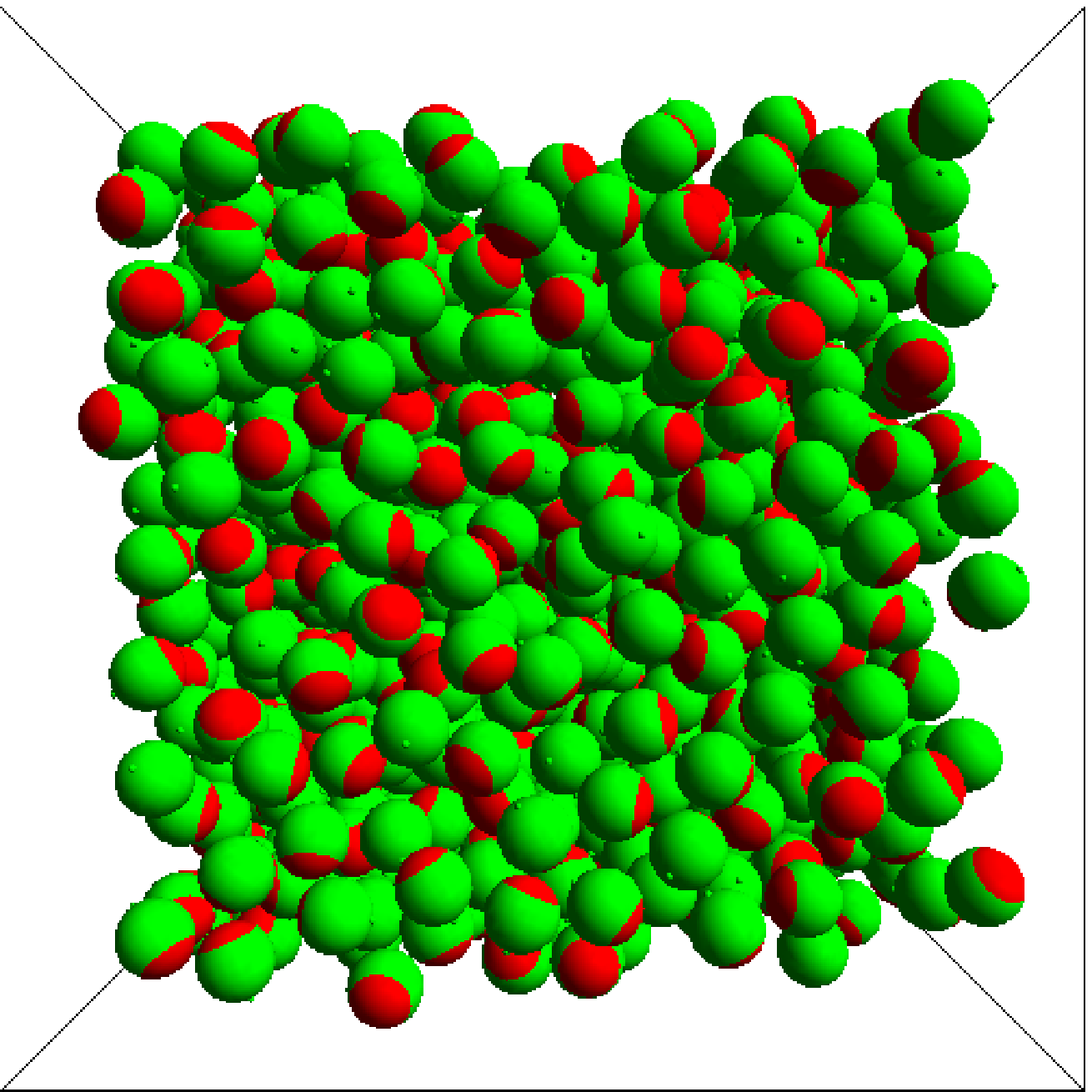} \\
\vskip0.5cm
\includegraphics[width=6cm,angle=0]{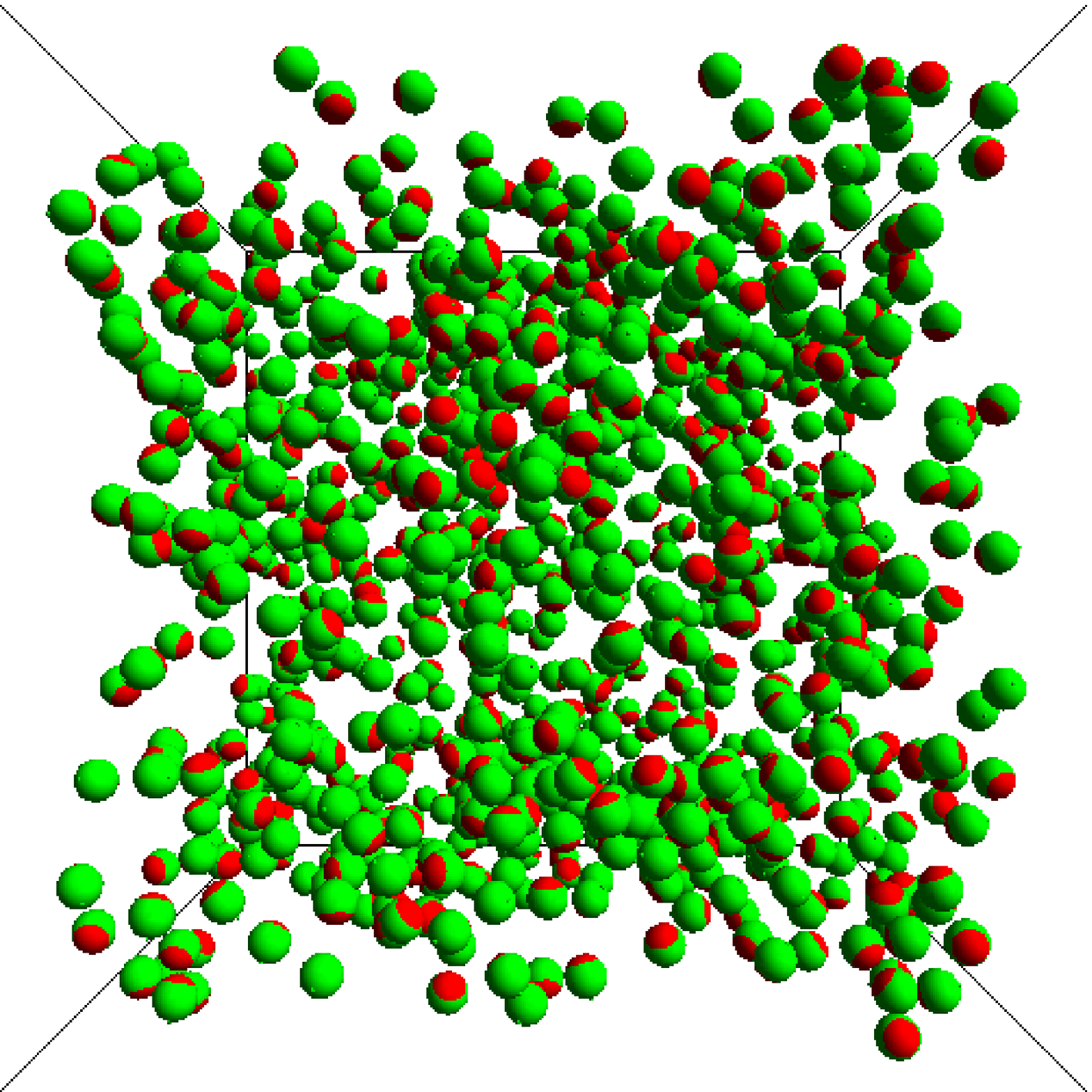} 
\caption{Plots of $\bar{g}(r,\cos \theta_{12})$ as a function of $r/\sigma$ for $\rho^{*}=0.68$ and temperatures $T^{*}=0.65$ (top panel) 
and $T^{*}=1.0$ 
(center panel). All results pertain to $\chi=0.8$ coverage and both MC and RHNC results are reported. The bottom panel depicts the same quantities for the 
low-density case $\rho^{*}=0.1$,  $T^{*}=1.0$. 
\label{fig:fig5}}
\end{figure}

Since the integral equation results provide a complete, albeit approximate, picture it is possible to calculate from them the simulation average of $g(12)$ over all orientations of 
$\hat{\mathbf{r}}_{12}(\Omega)$, while holding $\hat{\mathbf{n}}_{1}$, $\hat{\mathbf{n}}_{2}$, and $r_{12}$ fixed. This is detailed in Appendix \ref{app:pair}. The result is a quantity we denote $\bar{g}(r,\cos \theta_{12})$, which can be directly compared with the Monte Carlo results; this is done in Fig.~\ref{fig:fig5}. An overall agreement between  MC and RHNC results can be seen, although the RHNC performance clearly worsens as temperature decreases. It is worth emphasizing that, upon averaging, even an HT configuration shows a jump at $r/\sigma=1.5$, an indication of bonding in \textit{some} orientation of $\hat{\mathbf{r}}_{12}(\Omega)$. 

\begin{figure}[htbp]
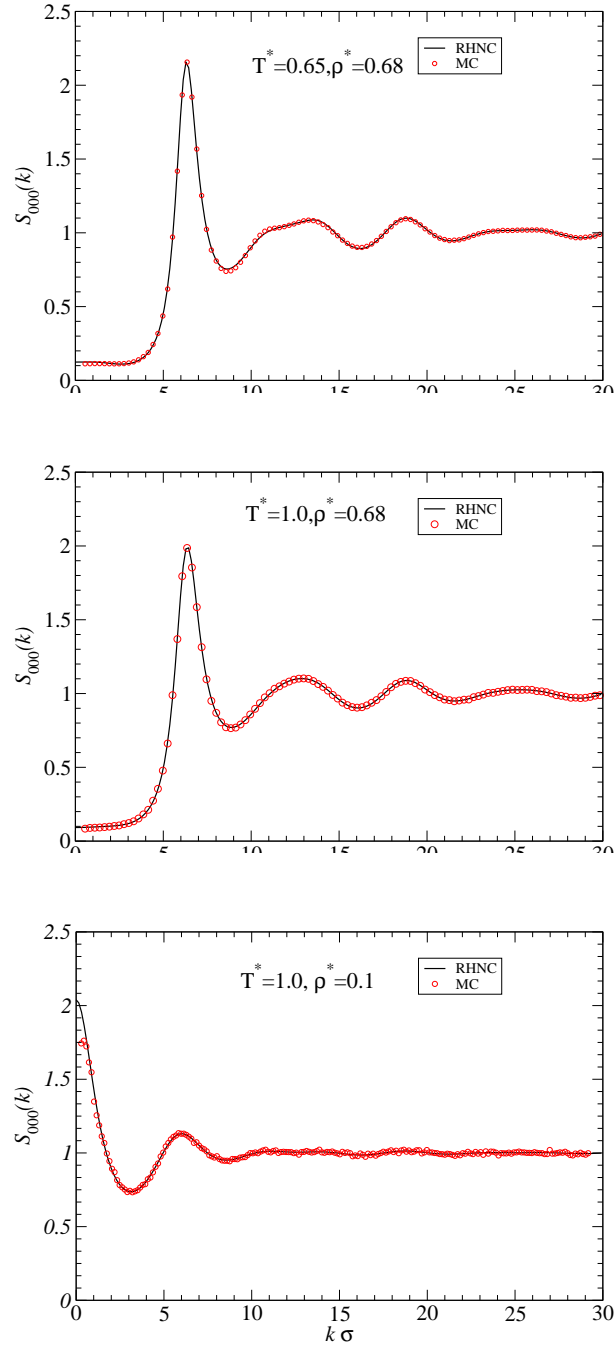

\includegraphics[width=8cm,angle=0]{Fig7a.eps}\\
\vskip0.5cm
\includegraphics[width=8cm,angle=0]{Fig7b.eps} \\
\vskip0.5cm
\includegraphics[width=8cm,angle=0]{Fig7c.eps} 
\caption{Plots of the coefficieunt $S_{000}(k)$ as a function of $k \sigma$ for $\rho^{*}=0.68$ and temperatures $T^{*}=0.65$ (top panel) and 
$T^{*}=1.0$ (center panel). 
The bottom panel depicts the low-density case $\rho^{*}=0.1$, $T^{*}=1.0$.
All results pertain to $\chi=0.8$ coverage and both MC and RHNC results are reported.
\label{fig:fig6}}
\end{figure}

Additional information about structural properties can be obtained from structure factors. In the presence of angular dependence, the quantity playing the role of the usual center-to-center structure factor is the $S_{000}(k)$ coefficient 
(see Appendix \ref{app:rhnc}) of $S(12) = 1+\rho \tilde{h}(12)$, which is plotted in Fig. \ref{fig:fig6} as a function of $k \sigma$ for the same state points as above. This spherical component of the structure is rather well predicted by the theory. Viewing the temperature dependence of $S_{000}(k)$, one notices that on cooling the second peak splits into two distinct peaks. 
Such a splitting, less evident but present also in the fourth peak, 
can be traced back to interference between the  asymptotic
structure factor oscillations due to the two discontinuities of the pair correlation functions at $\sigma$ and $\lambda \sigma$. By decreasing temperature, the
contribution of the pair correlation function close to $\lambda \sigma$ becomes
comparable with that from the region close to the hard sphere contact.
\subsection{Pair correlation functions and structure factor at different coverages}
\label{sec:pair_different}

As coverage decreases, it becomes harder and harder for integral equations to follow the critical line. This is because a lower coverage corresponds to a lower overall attractive interaction. As such, the critical temperatures and densities
progressively decrease and integral equations approximations and algorithms become less reliable.

Yet we can obtain interesting guidelines of the general trend by considering structural properties at different coverages. We then select a specific state point $\rho^{*}=0.68$ and $T^{*}=1.0$, which is above the critical temperatures for all coverages but still displays significant structural behavior.
Table \ref{tab:therm} reports the RHNC results for the main reduced thermodynamic quantities previously defined: the internal energy $\beta U/N$ per molecule,
the excess free energy per molecule $\beta F_{\text{ex}}/N$, the chemical potential $\beta \mu$ and the compressibility factor $\beta P/\rho$. In addition
we have also calculated the quantity $\beta (\partial P/\partial \rho)$, whose approach to zero signals the vicinity of a spinodal line. 

\begin{figure}[htbp]
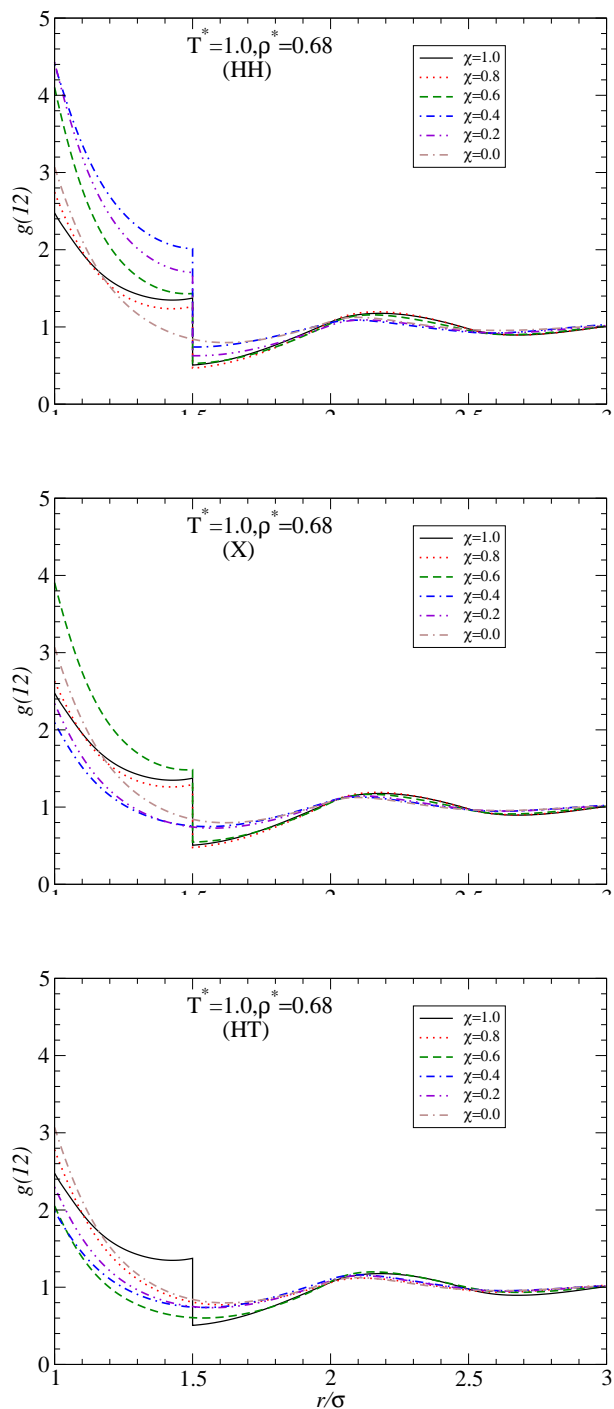

\includegraphics[width=8cm,angle=0]{Fig8a.eps}\\
\vskip0.5cm
\includegraphics[width=8cm,angle=0]{Fig8b.eps}\\
\vskip0.5cm
\includegraphics[width=8cm,angle=0]{Fig8c.eps} 
\caption{Plots of $g(12)$ as a function of $r/\sigma$ for $\rho^{*}=0.68$ and temperature $T^{*}=1.00$ at different coverages ranging from $\chi=1.0$ 
(fully symmetric square-well) to $\chi=0$ (hard spheres). 
Different panes refer to different configurations: HH (top panel), X (center panel), HT (bottom panel).   
Note that all curves have been drawn on the same scale. 
\label{fig:fig7}}
\end{figure}

Figure \ref{fig:fig7} again depicts $g(12)$ in the HH (top panel), X (central panel), and HT (bottom panel) configurations as a function of the coverage starting from a fully square-well potential ($\chi=1.0$) to a hard-sphere potential ($\chi=0.0$). The non-monotonic  dependence on coverage at the SW discontinuities $r=\sigma^{+}$ and $r=\lambda \sigma^{-}$ is particularly noteworthy and has been singled out in Table \ref{tab:g}. For brevity, we have put there $g_{\alpha} (r) \equiv g(r,\omega_1,\omega_2)$ for $r=\sigma^{+}$ and $r=\lambda\sigma^{-}$, where $\alpha$ stands for HH, X, or HT. 
Contact values within the well follow a rather erratic pattern as detailed in Table \ref{tab:g} and Fig. \ref{fig:fig7}.
The contact value $g_{\text{HH}}(\sigma^{+})$ first increases as $\chi$ decreases from $1$ to $0.5$ and then oscillates until a final sharp decrease to the hard-sphere limit at $\chi=0.0$. The particular high value of $\chi=0.1$ (see Table \ref{tab:g}) is a clear indication of a chaining and micelles formation,
as one expects for such a low coverage, but similar -- albeit lower --- values are found for $\chi=0.2$ and $\chi=0.4$, always below the $50\%$ limit. An even more irregular behavior can be observed at the other extreme of the well,
$g_{\text{HH}}(\lambda \sigma^{-})$.

\begin{figure}[htbp]
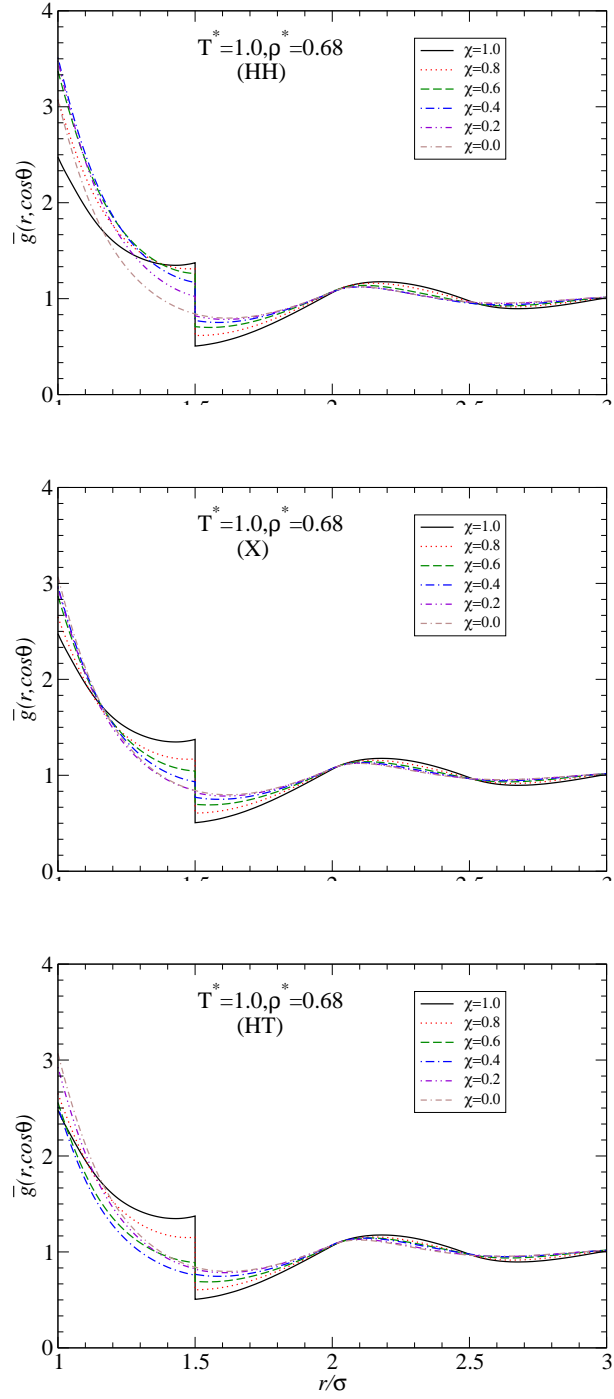

\includegraphics[width=8cm,angle=0]{Fig9a.eps}\\
\vskip0.5cm
\includegraphics[width=8cm,angle=0]{Fig9b.eps}\\
\vskip0.5cm
\includegraphics[width=8cm,angle=0]{Fig9c.eps} 
\caption{Plots of $\bar{g}(r,\cos \theta)$ as a function of $r/\sigma$ for $\rho^{*}=0.68$ and temperature $T^{*}=1.00$ at different coverages 
ranging from $\chi=1.0$ (fully symmetric square-well) to $\chi=0$ (hard spheres). Different panels refer to different configurations as before: HH (top panel), 
X (center panel), HT (bottom panel). Again, all curves have been drawn on the same scale. 
\label{fig:fig8}}
\end{figure}

This rich complexity cannot be explained by simple geometrical arguments hinging upon rotationally-averaged quantities. This can be quickly inferred by inspection of Table \ref{tab:gave}, where we report the average correlation function $\bar{g}(r,\cos \theta)$ defined as in the previous subsection, at the same interior well extremes $\sigma^{+}$ and $\lambda \sigma^{-}$. Figure \ref{fig:fig8} also depicts the coverage dependence of $\bar{g}(r,-1)$, $\bar{g}(r,0)$, and $\bar{g}(r,+1)$, respectively HH, X, and HT. Clearly, previous irregular behavior in the $g_{\alpha}(r)$, $\alpha=$ \{HH,X,HT\}, have been smoothed out and a much more regular trend is observed. For instance (see second column of Table \ref{tab:gave}), $\bar{g}(\sigma^{+},-1)$ gradually increases as $\chi$ decreases from 1.0, peaks at $0.3$, and drops to the hard-sphere limit. This non-monotonic behavior cannot be explained by simple heuristic arguments which would suggest a monotonic trend. 
As $\chi$ decreases, the whole coexistence curve is lowered on the temperature axis, effectively corresponding to an overall decrease of the attractive forces or, equivalently, to an increase of the temperature scale. 
This is consistent with a decrease of the contact value of the full coverage (square-well) contact value (see Fig.\ref{fig:fig3} top and central panel). Similar reasoning explains also the opposite trend followed by $\bar{g}(\lambda \sigma^{-},-1)$ (third column in Table \ref{tab:gave}, again contrasted with the square-well $\lambda \sigma^{-}$ behavior in Fig. \ref{fig:fig3} top and central panel, which increases as temperature decreases).

The perpendicular X configuration follows a similar pattern (see Fig. \ref{fig:fig7} and fourth and fifth columns in Tables \ref{tab:g} and \ref{tab:gave}) but has a sharp change in the $g(12)$ behavior upon passing from $\chi>0.5$ to $\chi<0.5$ (see Table \ref{tab:g}) indicating that $50\%$ coverage plays a special role, as one could expect from the the outset. Again these nuances are lost upon rotational averaging (see Table \ref{tab:gave}).

Next, we note that HT $g(12)$ configuration does not display any discontinuity at the well edge $\lambda \sigma^{-}$ 
(see bottom panel of Fig. \ref{fig:fig7})
as already remarked. 
Nevertheless, there is a significant coverage dependence interpolating between square-well and hard-sphere behavior again in a non-monotonic fashion 
which is smoothed out in corresponding rotational averaged quantity (see Fig.\ref{fig:fig8} and last two columns in Tables \ref{tab:g} and \ref{tab:gave}).

\begin{figure}[htbp]
\vskip0.5cm
\includegraphics[width=10cm,angle=0]{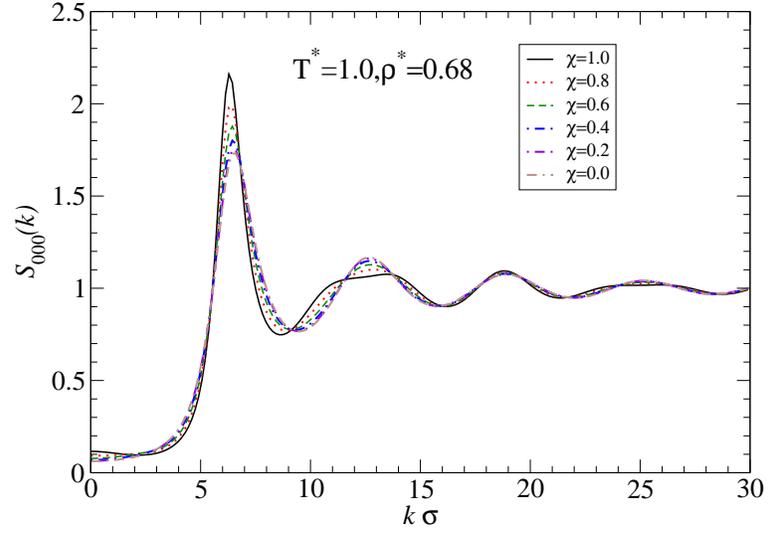}
\caption{Plots of $S_{000}(k)$ as a function of $k\sigma$ for $\rho^{*}=0.68$ and temperature $T^{*}=1.00$ at different coverages ranging 
from $\chi=1.0$ (fully symmetric square-well) to $\chi=0$ (hard spheres).
\label{fig:fig9}}
\end{figure}

As a final point, it is instructive to consider the coverage dependence of the radial structure factor $S_{000}(k)$. This is reported in Fig. \ref{fig:fig9}. As coverage decreases $S_{000}(0)$ shows a monotonic decrease, in agreement with the fact that the critical region is shifting to lower temperatures. The first peak is slightly shifted to higher $k\sigma $ values and decreases as coverage decreases. 
The second peak displays an opposite variation with the coverage,
accompanied by an increasing 
width. Such behavior can be understood as due to the increasing 
contribution of the discontinuity at $\lambda \sigma$ with increasing $\chi$ 
(see discussion of the interference of the asymptotic oscillations of
the structure factors at the end of Section V D).

\section{Conclusions and outlook}
\label{sec:conclusions}

We have studied the Kern-Frenkel model for hard spheres with a single attractive patch using integral equation techniques within the RHNC approximation and have compared results with various specialized Monte Carlo simulations. This potential constitutes a serious challenge to 
integral equation approaches for several reasons. First, because of the complexity in the underlying theory, which far exceeds that of its symmetrical counterpart. This is mirrored in the increased complexity of the corresponding numerical algorithms (see Appendix \ref{app:rhnc}). 
Second, the discontinuous nature of the patchy anisotropy poses in general a serious problem for theories based on expansions in spherical harmonics, as an infinite number of terms is in principle required to describe the discontinuity. While the first point is common to all molecular fluids, this second point is specific to patchy potentials. Remarkably, in the Kern-Frenkel potential case, it is possible to cope with this in a controlled manner by simply adapting the angular grid as detailed in the text. Finally, care is required when comparing pair correlation functions with Monte Carlo simulations, as different types of averages may be employed in simulations. Yet, since integral equations provide essentially complete descriptions within a given approximation defined by the closure, the pair correlation averages typically computed in MC simulations can be conveniently extracted from their results (see Appendix \ref{app:pair}).  

The RHNC integral equation has already been used in the past in the framework of associating  molecular fluids. \cite{Lado95,Lombardero99} It provides a robust and reliable description for a number of anisotropic potentials such as multipolar dependence. A major advantage of this particular closure, rendering RHNC particularly suited for computation of phase diagrams, is that the chemical potential calculation requires no additional approximation, thus enhancing internal thermodynamic consistency. This contrasts strongly with all other standard integral equation approaches (aside from the parent HNC), where pressure and chemical potential involve different approximations. Given a plausible representation of the bridge function, everything else follows, thus providing much better control over the procedures used.

We have given a full description of the phase diagram and structural properties of the one-patch model with a specific coverage ($80\%$). Notwithstanding a small systematic deviation in the detailed values, molecular RHNC correctly reproduces the general quantitative features of the phase diagram with an accuracy comparable to the analogous calculation of its spherical counterpart -- a simple square-well potential.

The pair correlation functions have been calculated for representative state points in the phase diagrams for three specific relative orientations of the patches on the pair spheres: when the two patches are aligned along the same direction (head-to-tail configurations), when they form a right angle (crossed configuration), and when they face each other (head-to-head configuration). The jumps characteristic of the square-well potential, the contact values, and the depth and position of the oscillations, all can be put in direct and natural correlation with the peculiarities of the patch dependence. Likewise, the $k=0$ value of the structure factor $S_{000}(k)$ and the position and morphology of the principal peaks can be directly related to the aggregation properties of the fluid and an appropriate angular average of the pair correlation function is amenable to a direct comparison with Monte Carlo simulations. We have shown that, even in the fluid-like regime, RHNC provides a rather precise description.

Finally, we have provided a detailed study of the pair correlation functions and structure factor for different values of the coverage (or, equivalently, for different size of the patch). We find a flourish of different behaviors, with a non-monotonic dependence on the coverage, which are inaccessible to numerical simulations based on rotationally averaged quantities and could help to interpret the low-coverage fate of the coexistence line.

The present RHNC study has two main limitations. First we observe a significant, albeit small, quantitative difference with numerical simulations in the prediction of the critical line location. This is due on the one hand to intrinsic drawbacks of the RHNC approximations which are present also in the fully isotropic spherical case, \cite{Giacometti09} and, on the other hand, to the extremely anisotropic nature of the single-patch potential, which emphasizes the shortcomings of the isotropic approximation used for the bridge function in the present work. Secondly we have restricted to a single representative coverage ($80\%$) for the coexistence line location. This is because of the numerical instabilities arising for lower coverages in view of the much lower temperatures required. Both those shortcomings will be addressed in a future publication.  

A number of possible perspectives can be envisaged from the results of the present paper. As the coverage decreases, the binodal curve shrinks and moves to the left, thus rendering integral equation algorithms less and less stable, as remarked. This is nevertheless an extremely interesting region to study with integral equations, due to the subtle interplay between condensation and polymerization which could give rise to a very rich phenomenology. \cite{Dudo_04,VanWorkum05,Teixeira09} In addition, a comparison with the case of two patches, set on opposite sides of a sphere, at the same global coverage could illuminate the critical-point dependence on both valence and size distribution, a question of experimental relevance. Preliminary results for the two-patches case clearly show that indeed RHNC is able to closely follow numerical simulation predictions for the critical line at different coverages. A detailed analysis of this and other related points will be reported elsewhere.

\section{Acknowledgments}
FS acknowledges support from  NoE SoftComp NMP3-CT-2004-502235 and ERC-226207-PATCHYCOLLOIDS.
AG acknowledges support from PRIN-COFIN 2008/2009. 

\appendix

\section{Key expressions for the integral equation procedure}
\label{app:rhnc}
For the sake of compactness and future reference, we provide below the main expressions necessary for the numerical solution of OZ-type integral equations with orientational dependence. The resulting iteration scheme is laid out in Table I.

\subsection{Ornstein-Zernike equation and closure}
\label{app:rhnc_a}

The OZ equation in terms of $\gamma(12)=h(12)-c(12)$ is
\begin{eqnarray}
\label{app:eq1}
\gamma\left(12\right) &=& \frac{\rho}{4\pi} \int d\mathbf{r}_3 d\omega_3 \left[ \gamma\left(13\right)+c\left(13\right)\right] 
c\left(32\right),
\end{eqnarray}
while the general form of a closure is
\begin{eqnarray}
\label{app:eq2}
c\left(12\right)&=& \exp \left[-\beta \Phi\left(12\right)+\gamma\left(12\right)+
B\left(12\right)\right]-1-\gamma\left(12\right).
\end{eqnarray}
The bridge function $B\left(12\right)$ is known only as an infinite power series in density whose coefficients cannot be readily calculated. \cite{Hansen86} All practical closures approximate $B\left(12\right)$ in some way.

\subsection{Expansions in spherical harmonics}
\label{app:rhnc_b}

The expansion in spherical harmonics $Y_{lm}(\omega)$ of any correlation function
$X(12)$ in its axial frame ($\hat{\mathbf{z}} = \hat{\mathbf{r}}$) is given by
\begin{eqnarray}
\label{app:eq3}
X(12) &=& X\left(r,\omega_1,\omega_2 \right) = 4\pi \sum_{l_1,l_2,m} X_{l_1 l_2 m} 
\left(r\right) Y_{l_1m}\left(\omega_1\right) Y_{l_2\bar{m}}\left(\omega_2\right),
\end{eqnarray}
where $\bar{m} \equiv -m$, and its inverse reads
\begin{eqnarray}
\label{app:eq4}
X_{l_{1} l_{2} m} \left(r\right)&=&\frac{1}{4\pi} \int d\omega_1d\omega_2 X\left(r,\omega_1,\omega_2 \right) 
Y_{l_{1}m}^*\left(\omega_1\right)
Y_{l_{2}\bar{m}}^*\left(\omega_2\right). 
\end{eqnarray}
The same expressions, with $k$ replacing $r$, apply to the Fourier transform $\tilde{X}(12)$ in momentum space, where the axial frame is now defined by 
$\hat{\mathbf{z}} = \hat{\mathbf{k}}$. The coefficients in the expansions satisfy the symmetries $X_{l_{1}l_{2}\bar{m}}=X_{l_{1}l_{2}m}$ 
and $X_{l_{2}l_{1}m}=(-1)^{l_1+l_2}X_{l_{1}l_{2}m}$.

The angular integrations in the expansion inverse (\ref{app:eq4}) are evaluated using Gaussian quadratures. Note that to compute $c_{l_{1} l_{2} m} \left(r\right)$ 
from the closure equation (\ref{app:eq2}) using (\ref{app:eq4}), we must first construct $\gamma(12)$ and (an approximate) $B(12)$ from their axial expansion coefficients using (\ref{app:eq3}); the potential $\Phi\left(12\right)$ is already in angular representation.

The OZ equation factors in momentum space; expressed in terms of the axial expansion coefficients of the transformed pair functions, we get the simple matrix form
\begin{eqnarray}
\label{app:eq5}
\tilde{\gamma}_{l_{1} l_{2} m}\left(k\right) &=& \left(-1\right)^{m} \rho
\sum_{l_{3}=m}^{\infty} 
\left[\tilde{\gamma}_{l_{1} l_{3} m}\left(k\right)+
\tilde{c}_{l_{1} l_{3} m}\left(k\right) \right]
\tilde{c}_{l_{3} l_{2} m}\left(k\right),
\end{eqnarray}
which can be solved for the $\tilde{\gamma}_{l_{1} l_{2} m}\left(k\right)$ using standard matrix operations. (A carryover from the Hankel and Clebsch-Gordan transforms defined below is that transform coefficients with $l_1+l_2$ even are real while for $l_1+l_2$ odd they are imaginary. Matrices such as 
$\tilde{\mathbf{c}}_m(k)$ with elements $\tilde{c}_{l_1 l_2 m}(k)$, $l_1,l_2 \ge m$, are Hermitian.)

\subsection{Clebsch-Gordan transforms}
\label{app:rhnc_c}

For the function $X(12)$ in direct space, the Clebsch-Gordan transform from axial coefficients $X_{l_{1} l_{2} m}(r)$ when 
$\hat{\mathbf{z}} = \hat{\mathbf{r}}$ to space coefficients $X\left(r;l_{1} l_{2} l \right)$ is
\begin{eqnarray}
\label{app:eq6}
X\left(r;l_{1} l_{2} l \right)&=& \left(\frac{4 \pi}{2l+1}\right)^{1/2}
\sum_{m} C \left(l_{1} l_{2} l ; m \bar{m} 0 \right) X_{l_{1} l_{2} m} \left(r\right),
\end{eqnarray}
where the $C \left(l_{1} l_{2} l ; m \bar{m} 0 \right)$ are Clebsch-Gordan coefficients;
the inverse transform is given by 
\begin{eqnarray}
\label{app:eq7}
X_{l_{1} l_{2} m} \left(r\right)&=& \sum_{l} C \left(l_{1} l_{2} l ; m \bar{m} 0 \right) \left(\frac{2l+1}{4\pi}\right)^{1/2} 
X\left(r;l_{1} l_{2} l\right).
\end{eqnarray}
The same Clebsch-Gordan transforms, with $k$ replacing $r$, apply to the Fourier transform $\tilde{X}(12)$ in momentum space, where the axial frame is now defined by $\hat{\mathbf{z}} = \hat{\mathbf{k}}$. Symmetry requires that $l_1+l_2+l$ be an even integer.

\subsection{Hankel transforms}
\label{app:rhnc_d}

The Fourier transform in an arbitrary space frame of a function $X(12)$ into $\tilde{X}(12)$ leads to the Hankel transforms of the coefficients, 
\begin{eqnarray}
\label{app:eq8}
\tilde{X}\left(k;l_{1} l_{2} l \right) &=& 4 \pi \mathrm{i}^{l}
\int_{0}^{\infty} dr~ r^2 X\left(r;l_{1} l_{2} l \right) j_{l} \left(k r\right),
\end{eqnarray}
with the inverse transform reading
\begin{eqnarray}
\label{app:eq9}
X\left(r;l_{1} l_{2} l \right) &=&
\frac{1}{2 \pi^2 \mathrm{i}^l} \int_{0}^{\infty} dk~ k^2 
\tilde{X}\left(k;l_{1} l_{2} l  \right) j_l \left(kr\right),
\end{eqnarray}
where $j_l(x)$ is the spherical Bessell function of order $l$. By using ``raising'' and ``lowering'' operations on the integrands, \cite{Lado82a} these are finally 
evaluated with $j_0(kr)=\sin kr/kr$ kernels, for $l$ even, and $j_{-1}(kr)=\cos kr/kr$ kernels, for $l$ odd. Fast Fourier Transforms are programmed for both instances. (Note that Hankel transforms $\tilde{X}\left(k;l_{1} l_{2} l \right)$ are imaginary for $l$ odd while for $l$ even they are real.)

\section{Comparing integral equation pair correlation functions with numerical simulation results}
\label{app:pair}

The function $X(12)$, expanded in its axial frame where $\hat{\mathbf{z}} = \hat{\mathbf{r}}$, is given in Eq. (\ref{app:eq3}). If we rotate the frame of reference into an arbitrary space frame, its spherical harmonic expansion becomes \cite{Lado82a}
\begin{eqnarray}
\label{pair:eq1}
X\left(12\right) &=& 4 \pi \sum_{l_{1},l_{2},l} X\left(r;l_{1} l_{2} l\right)
\sum_{m_{1},m_{2}} C\left(l_{1} l_{2} l; m_{1} m_{2} m_{1}+m_{2} \right)
Y_{l_{1}m_{1}}\left(\omega_{1}\right) Y_{l_{2}m_{2}}\left(\omega_{2}\right)
Y_{l, m_{1}+m_{2}}^{*}\left(\Omega\right),
\end{eqnarray}
where the $X\left(r;l_{1} l_{2} l\right)$ are related to the original $X_{l_{1} l_{2} m}(r)$ through the Clebsch-Gordan transform Eq. (\ref{app:eq6}). In order to compare correlation functions such as $g(12)$ with the results of our Monte Carlo simulations, we need to average $g(12)$ written as in Eq. (\ref{pair:eq1}) over all possible orientations of $\hat{\mathbf{r}}(\Omega)$, keeping 
$\hat{\mathbf{n}}_{1}(\omega_{1})$, $\hat{\mathbf{n}}_{2}(\omega_{2})$, and $r$ \textit{fixed}. By symmetry, the result can depend only on $r$ and 
$\cos\theta_{12} \equiv \hat{\mathbf{n}}_1 \cdot \hat{\mathbf{n}}_2$. We define thus
\begin{eqnarray}
\label{pair:eq2}
\bar{g}\left(r,\cos\theta_{12}\right) &=& \frac{1}{4\pi} \int d\Omega~ g(12). \end{eqnarray}
We find first that
\begin{eqnarray}
\label{pair:eq3}
\frac{1}{4\pi} \int d\Omega~ Y_{l, m_{1}+m_{2}}^{*}\left(\Omega\right) &=&  
\frac{1}{(4\pi)^{1/2}} \int d\Omega~ Y_{00}\left(\Omega\right)Y_{l,m_{1}+m_{2}}^{*}\left(\Omega\right)
= \frac{\delta_{l0} \delta_{m_{1}+m_{2},0}}{(4\pi)^{1/2}},
\end{eqnarray}
so that Eq. (\ref{pair:eq2}) becomes, with (\ref{pair:eq1}),
\begin{eqnarray}
\label{pair:eq4}
\bar{g}\left(r,\cos\theta_{12}\right) &=& (4\pi)^{1/2} \sum_{l_{1},l_{2},m} g\left(r;l_{1} l_{2} 0\right) 
C\left(l_{1} l_{2} 0; m \bar{m} 0 \right)
Y_{l_{1}m}\left(\omega_{1}\right) Y_{l_{2}\bar{m}}\left(\omega_{2}\right).
\end{eqnarray}
Then using \cite{Gray84} 
\begin{eqnarray}
\label{pair:eq5}
C\left(l_{1} l_{2} 0;m \bar{m} 0\right) &=& \frac{\left(-1\right)^{l_1+m}}{(2 l_{1}+1)^{1/2}}~ \delta_{l_{1} l_{2}}
\end{eqnarray}
and the addition theorem for spherical harmonics, \cite{Gray84} one finds finally that
\begin{eqnarray}
\label{pair:eq6}
\bar{g}\left(r,\cos\theta_{12}\right) &=& \sum_{l=0}^\infty \left(-1\right)^l \left( \frac{2l+1}{4\pi} \right)^{1/2} g\left(r;ll0 \right) 
P_{l}\left(\cos \theta_{12}\right),
\end{eqnarray}
where $P_l(\cos \theta)$ is the usual Legendre polynomial of order $l$.


\begin{thebibliography}{99}

\bibitem{Wertheim84}
M.S. Wertheim, J. Stat. Phys. \textbf{35}, 19 (1984).

\bibitem{Nezbeda05} 
I.~Nezbeda, Mol. Phys. \textbf{103}, 59 (2005).

\bibitem{Carlevaro03}
C.M. Carlevaro, L. Blum, and F. Vericat, J. Chem. Phys. \textbf{119}, 5198 (2003).

\bibitem{Varakin97}
E. Vakarin, Y. Duda, and M.F. Holovko, Mol. Phys. \textbf{90}, 611 (1997). 

\bibitem{Jirsak07}
J. Jirs{\' a}k and I. Nezbeda, J. Chem. Phys. \textbf{127}, 124508 (2007).

\bibitem{Glotzer07} S.C. Glotzer and M.J. Solomon, Nature Materials \textbf{6}, 557 (2007).

\bibitem{Lomakin99}
A.~Lomakin, N.~Asherie, and G.B. Benedek., Proc. Natl. Acad. Sci. \textbf{96}, 9465 (1999).

\bibitem{McManus07}
J.J. McManus, A.~Lomakin, O.~Ogun, A.~Pande, M.~Basan, J.~Pande, and G.B. Benedek,
Proc. Natl. Acad. Sci. \textbf{104}, 16856 (2007).

\bibitem{Liu07}
H.~Liu, S.K. Kumar, and F.~Sciortino, J. Chem. Phys. \textbf{127}, 084902 (2007).

\bibitem{Pellicane08} 
G. Pellicane, G. Smith, and L. Sarkisov, Phys. Rev. Lett. \textbf{101}, 248102 (2008).

\bibitem{Teixeira00}
P.I.C. Teixeira, J.M. Tavares, and M.M. Telo da Gama, J. Phys.: Cond. Matt. \textbf{12}, R411 (2000).

\bibitem{Horsch05}
M.A. Horsch, Z. Zhang, and S.C. Glotzer, Phys. Rev. Lett. \textbf{95}, 056105 (2005).

\bibitem{Bianchi06} 
E.~Bianchi, J.~Largo, P.~Tartaglia, E.~Zaccarelli, and F.~Sciortino, Phys. Rev. Lett. \textbf{97}, 168301 (2006).

\bibitem{Foffi07} 
G. Foffi and F. Sciortino, J. Phys. Chem. B \textbf{111}, 9702 (2007).

\bibitem{Bianchi07}
E. Bianchi, P. Tartaglia, E. La Nave, and F. Sciortino, J. Phys. Chem. B \textbf{111}, 11765 (2007).  

\bibitem{Fantoni07}
R.~Fantoni, D. Gazzillo, A.~Giacometti, M.A. Miller, and G.~Pastore,
J. Chem. Phys. \textbf{127}, 234507 (2007).

\bibitem{Kern03} N. Kern and D. Frenkel, J. Chem. Phys. \textbf{118}, 9882 (2003).

\bibitem{Chapman88} W.G. Chapman, G. Jackson, and K.E. Gubbins, Mol. Phys. \textbf{65}, 1057 (1988).

\bibitem{Lado73} F. Lado, Phys. Rev. A \textbf{8}, 2548 (1973).

\bibitem{Lado82a} F. Lado, Mol. Phys. \textbf{47}, 283 (1982).

\bibitem{Lado82b} F. Lado, Mol. Phys. \textbf{47}, 299 (1982).

\bibitem{Goegelein08} C. G\"ogelein, G. N\"agele, R. Tuinier, T. Gibaud, A. Stradner, and P. Schurtenberger, 
J. Chem. Phys. \textbf{129}, 085102 (2008).

\bibitem{Kolafa87} J. Kolafa and I. Nezbeda, Mol. Phys.  \textbf{61}, 161 (1987).

\bibitem{Demichele1} C. De Michele, P. Tartaglia, and F. Sciortino
J. Chem. Phys. \textbf{125}, 204710 (2006). 

\bibitem{Demichele2} C. De Michele, S. Gabrielli, P. Tartaglia, and F. Sciortino
J. Phys. Chem. B \textbf{110}, 8064 (2006).

\bibitem{Russo09} J. Russo, P. Tartaglia, and F. Sciortino, J. Chem. Phys. \textbf{131}, 014504 (2009).

\bibitem{Rosenfeld79} Y. Rosenfeld and N.W. Ashcroft, Phys. Rev. A \textbf{20}, 1208 (1979).

\bibitem{comment} Rosenfeld and Ashcroft note that $\sigma_0$ could be determined by requiring consistency between the virial and compressibility equations of state but do not implement this route in Ref. \onlinecite{Rosenfeld79}.

\bibitem{Lado82} F. Lado, Phys. Lett. \textbf{89}A, 196 (1982).

\bibitem{Hiroike57} K. Hiroike, J. Phys. Soc. Japan \textbf{12}, 326 (1957).

\bibitem{Lado95} F. Lado, E. Lomba, and M. Lombardero, J. Chem. Phys. \textbf{103}, 481 (1995).

\bibitem{Lombardero99} M. Lombardero, C. Mart\'{\i}n, S. Jorge, F. Lado, and E. Lomba, J. Chem. Phys. \textbf{110}, 1148 (1999).

\bibitem{Gray84} C.G. Gray and K.E. Gubbins, \textit{Theory of Molecular Fluids. Vol 1: Fundamentals} (Clarendon, Oxford, 1984). 

\bibitem{Morita60} T. Morita and K. Hiroike, Prog. Theor. Phys. \textbf{23}, 1003 (1960).

\bibitem{Frenkel02} B. Smith and D. Frenkel, \textit{Understanding Molecular Simulation: From Algorithms to Applications} (Academic, San Diego, 
2002).

\bibitem{Wilding97} 
N.B. Wilding, J. Phys.: Cond. Matt. \textbf{9}, 585 (1997).
 
\bibitem{Hansen86} J.P. Hansen and I.R. McDonald, \textit{Theory of Simple Liquids} (Academic, New York, 1986).

\bibitem{Giacometti09} A. Giacometti, G. Pastore, and F. Lado, Mol. Phys. \textbf{107}, 555 (2009).

\bibitem{Verlet72} L. Verlet and J.J. Weis, Phys. Rev. A \textbf{5}, 939 (1972).

\bibitem{Henderson75} D. Henderson and E.W. Grundke, J. Chem. Phys. \textbf{63}, 601 (1975).

\bibitem{Schlijper93} A.G. Schlijper, M.M. Telo da Gama, and P.G. Ferreira, J. Chem. Phys. \textbf{98}, 1534 (1993).

\bibitem{Charpentier05} I. Charpentier and N. Jakse, J. Chem. Phys. \textbf{123}, 204910 (2005).

\bibitem{ElMendoub08} E.B. El Mendoub, J.F. Wax, I. Charpentier, and N. Jakse, Mol. Phys. \textbf{106}, 2667 (2008).

\bibitem{Belloni93} L. Belloni, J. Chem. Phys. \textbf{98}, 8080 (1993).

\bibitem{Dudo_04} J. Dudowicz, K.F. Freed, and J.F. Douglas,
Phys. Rev. Lett. \textbf{92}, 045502 (2004).

\bibitem{VanWorkum05} K. Van Workum  and J. F. Douglas, Phys. Rev. E \textbf{71} 031502 (2005).

\bibitem{Teixeira09} J.M. Tavares, P.I.C. Teixeira, and M.M. Telo da Gama, Mol. Phys. \textbf{107}, 453 (2009).

\end{thebibliography}

\clearpage
\begin{table} 
\begin{equation*} 
\begin{CD}
c\left(r;l_{1}l_{2}l\right)
@>\text{Hankel transform Eq. (\ref{app:eq8})}>>
\tilde{c}\left(k;l_{1}l_{2}l\right)
@>\text{Inverse CG transform Eq. (\ref{app:eq7})}>>
\tilde{c}_{l_{1}l_{2}m}\left(k\right) \\
@AA\text{CG transform Eq. (\ref{app:eq6})} A  @.    @VV\text{OZ equation Eq. (\ref{app:eq5}) }V\\
c_{l_{1}l_{2}m}\left(r\right)
@.      @.
\tilde{\gamma}_{l_{1}l_{2}m}\left(k\right) \\     
@AA\text{Expansion inverse Eq. (\ref{app:eq4})} A  @.    @VV\text{CG transform  Eq. (\ref{app:eq6})}V\\
c\left(r,\omega_1,\omega_2\right)
@.      @.
\tilde{\gamma}\left(k;l_{1}l_{2}l\right) \\
@AA\text{Closure Eq. (\ref{app:eq2})} A  @.    @VV\text{Inverse Hankel transform  Eq. (\ref{app:eq9})}V\\
\gamma\left(r,\omega_1,\omega_2\right)   
@.      @.
\gamma\left(r;l_{1} l_{2} l\right) \\
@AA\text{Expansion Eq. (\ref{app:eq3})} A   @.   @VV\text{Inverse CG transform Eq. (\ref{app:eq7}) }V\\
\left[\gamma_{l_{1} l_{2} m}\left(r\right) \right]_{\text{old}}   
@<<<    \text{Iterate}    @<<<  
\left[\gamma_{l_{1} l_{2} m}\left(r\right) \right]_{\text{new}}
\end{CD}
\end{equation*}
\label{tab:tab1}
\caption{Schematic flow-chart for the solution of the OZ equation
for angle-dependent potentials.}
\end{table}

\clearpage
\begin{table}
\begin{center}
\begin{tabular}{ccccccc}
\hline
$\exp(\beta \mu)$ \text{(GEMC)} && $\rho^{*}$ \text{(GEMC)}&&$\exp(\beta \mu)$ \text{(RHNC)} &&$\rho^{*}$ \text{(RHNC)}\\
\hline
$0.0105$ && $0.6044$     &&$0.0119$  && $0.65$     \\
$0.0120$ && $0.6287$     &&$0.0165$  && $0.68$     \\
$0.0150$ && $0.6555$     &&$0.0219$  && $0.70$     \\
$0.0180$ && $0.6752$     &&$0.0380$  && $0.73$     \\
$0.0210$ && $0.6845$     &&$0.0596$  && $0.75$     \\
$0.0240$ && $0.6934$     &&$0.0997$  && $0.77$     \\
$-$      && $-$          &&$0.2408$  && $0.88$     \\
\hline
\end{tabular}
\caption[]{Tabulated values for $\exp(\beta \mu)$ and $\rho^{*}$ corresponding to the liquid branch of Figure \ref{fig:fig1} for both GEMC and RHNC data. Note that these values are results of different calculations in the GEMC and RHNC cases and the two abscissas are not the same. Here and below expected errors are in the last digits.    
\label{tab:expbetamu}
}
\end{center}
\end{table}

\clearpage
\begin{table}
\begin{center}
\begin{tabular}{lcccccccc}
\hline
$T^{*}$ && $\rho_g^{*}$ &&$\rho_l^{*}$ &&$\beta \mu$ && $\beta P$\\
\hline
$0.50$ && $0.0016$     &&$0.7465$    && $-6.4789$ && $0.0016$\\
$0.55$ && $0.0040$     &&$0.7173$    && $-5.6342$ && $0.0038$\\
$0.60$ && $0.0086$     &&$0.6870$    && $-4.9607$ && $0.0077$\\       
$0.65$ && $0.0171$     &&$0.6518$    && $-4.4147$ && $0.0141$\\        
$0.70$ && $0.0306$     &&$0.6098$    && $-3.9668$ && $0.0228$\\        
$0.75$ && $0.0414$     &&$0.5436$    && $-3.5965$ && $0.0340$\\       
\hline
\end{tabular}
\caption[]{Numerical values of the RHNC coexistence line of
Figure \ref{fig:fig2}. Here $\rho_g^{*}$ and $\rho_l^{*}$ are the reduced densities of the gas and liquid branches, $T^{*}$ is the corresponding coexistence reduced temperature, and $\beta \mu$ and $\beta P$ are the common values of the chemical potential and pressure.     
\label{tab:coextable}
}
\end{center}
\end{table}

\clearpage
\begin{table}
\begin{center}
\begin{tabular}{lccccccccccccc}
\hline
$\chi$ && $\frac{U}{N\epsilon}$&&$\frac{\beta F_{\text{ex}}}{N}$ &&$\beta \mu$&&$\frac{\beta P}{\rho}$
&& $\beta \left(\frac{\partial P}{\partial \rho}\right)$ && $\frac{\sigma_0}{\sigma}$\\
\hline
$1.0$ && $-5.32$     &&$-2.55$  && $-3.58$      &&$0.35$   &&  $ 8.62$  && $1.031$  \\
$0.9$ && $-4.52$     &&$-1.70$  && $-1.82$      &&$1.27$   &&  $ 9.87$  && $1.025$  \\
$0.8$ && $-3.76$     &&$-0.92$  && $-0.24$      &&$2.07$   &&  $10.86$  && $1.018$  \\
$0.7$ && $-3.02$     &&$-0.18$  && $ 1.24$      &&$2.81$   &&  $11.98$  && $1.014$  \\
$0.6$ && $-2.37$     &&$ 0.45$  && $ 2.49$      &&$3.43$   &&  $12.88$  && $1.010$  \\
$0.5$ && $-1.77$     &&$ 1.02$  && $ 3.60$      &&$3.96$   &&  $13.75$  && $1.007$  \\
$0.4$ && $-1.21$     &&$ 1.52$  && $ 4.57$      &&$4.44$   &&  $14.49$  && $1.005$  \\
$0.3$ && $-0.75$     &&$ 1.91$  && $ 5.32$      &&$4.80$   &&  $15.08$  && $1.003$  \\
$0.2$ && $-0.37$     &&$ 2.22$  && $ 5.92$      &&$5.09$   &&  $15.60$  && $1.002$  \\
$0.1$ && $-0.11$     &&$ 2.41$  && $ 6.33$      &&$5.30$   &&  $15.85$  && $1.000$  \\
$0.0$ && $ 0.00$     &&$ 2.49$  && $ 6.47$      &&$5.37$   &&  $15.99$  && $1.000$  \\
\hline
\end{tabular}
\caption[]{Values of reduced internal and excess free energies, chemical potential, pressure, and inverse compressibility $\beta \left(\frac{\partial P}{\partial \rho}\right)$ as a function of the coverage $\chi$ for a fixed state point $\rho^{*}=0.68$ and $T^{*}=1.0$. The last
column gives the corresponding values of $\sigma_0/\sigma$.
\label{tab:therm}
}
\end{center}
\end{table}

\begin{table}
\begin{center}
\begin{tabular}{lcccccccccccccc}
\hline
$\chi$ && $g_{\text{HH}}(\sigma^{+})$&&$g_{\text{HH}}(\lambda \sigma^{-})$ &&$g_{\text{X}}(\sigma^{+})$&&$g_{\text{X}}(\lambda \sigma^{-})$
&& $g_{\text{HT}}(\sigma^{+})$ && $g_{\text{HT}}(\lambda \sigma^{-})$&& $\bar{z}$ \\
\hline
$1.0$ && $2.476$   &&$1.367$&& $2.476$    &&$1.367$&&  $2.476$   & & $1.367$  && $10.64$\\
$0.9$ && $2.574$   &&$1.527$&& $2.499$    &&$1.468$&&  $2.449$   & & $0.966$  && $ 9.05$\\
$0.8$ && $2.745$   &&$1.257$&& $2.631$    &&$1.284$&&  $2.781$   & & $0.814$  && $ 7.53$\\
$0.7$ && $3.287$   &&$1.211$&& $3.092$    &&$1.280$&&  $2.367$   & & $0.649$  && $ 6.04$\\
$0.6$ && $4.101$   &&$1.431$&& $3.906$    &&$1.483$&&  $2.052$   & & $0.608$  && $ 4.75$\\
$0.5$ && $4.643$   &&$1.802$&& $4.907$    &&$1.806$&&  $1.941$   & & $0.659$  && $ 3.53$\\
$0.4$ && $4.402$   &&$2.015$&& $2.094$    &&$0.758$&&  $1.986$   & & $0.745$  && $ 2.42$\\
$0.3$ && $4.025$   &&$1.869$&& $2.219$    &&$0.771$&&  $2.093$   & & $0.778$  && $ 1.51$\\
$0.2$ && $4.429$   &&$1.708$&& $2.341$    &&$0.747$&&  $2.298$   & & $0.754$  && $ 0.74$\\
$0.1$ && $6.199$   &&$1.918$&& $2.678$    &&$0.776$&&  $2.668$   & & $0.773$  && $ 0.22$\\
$0.0$ && $3.069$   &&$0.849$&& $3.069$    &&$0.849$&&  $3.069$   & & $0.849$  && $ 0.0 $\\
\hline
\end{tabular}
\caption[]{Behavior of $g_{\alpha}(\sigma^{+})$ and $g_{\alpha}(\lambda \sigma^{-})$, 
where $\alpha=\{\text{HH,X,HT}\}$. As in the text, HH, X, and HT refer to $\hat{\mathbf{n}}_1 \cdot \hat{\mathbf{n}}_2=-1,0,1$ respectively. All values refer to the same state point $\rho^{*}=0.68$ and $T^{*}=1.0$. Here and below, $\sigma^{+}$ and $\lambda \sigma^{-}$ refer to values at discontinuities inside the well, $\sigma < r < \lambda \sigma$. The last column gives the average coordination number $\bar{z}$ within the well.
\label{tab:g}
}
\end{center}
\end{table}
%
\begin{table}
\begin{center}
\begin{tabular}{lccccccccccccc}
\hline
$\chi$ && $\bar{g}(\sigma^{+},-1)$&&$\bar{g}(\lambda \sigma^{-},-1)$ &&$\bar{g}(\sigma^{+},0)$&&$\bar{g}(\lambda \sigma^{-},0)$
&& $\bar{g}(\sigma^{+},+1)$ && $\bar{g}(\lambda \sigma^{-},+1)$ \\
\hline
$1.0$ && $2.476$   &&$1.367$&& $2.476$    &&$1.367$&&  $2.476$   & & $1.367$  \\
$0.9$ && $2.802$   &&$1.343$&& $2.585$    &&$1.264$&&  $2.591$   & & $1.263$  \\
$0.8$ && $3.073$   &&$1.312$&& $2.650$    &&$1.169$&&  $2.653$   & & $1.151$  \\
$0.7$ && $3.238$   &&$1.289$&& $2.762$    &&$1.101$&&  $2.681$   & & $1.032$  \\
$0.6$ && $3.373$   &&$1.263$&& $2.869$    &&$1.046$&&  $2.558$   & & $0.891$  \\
$0.5$ && $3.444$   &&$1.220$&& $2.927$    &&$0.989$&&  $2.407$   & & $0.775$  \\
$0.4$ && $3.513$   &&$1.174$&& $2.969$    &&$0.938$&&  $2.513$   & & $0.764$  \\
$0.3$ && $3.540$   &&$1.114$&& $2.977$    &&$0.891$&&  $2.773$   & & $0.810$  \\
$0.2$ && $3.480$   &&$1.031$&& $2.955$    &&$0.849$&&  $2.926$   & & $0.832$  \\
$0.1$ && $3.380$   &&$0.952$&& $3.016$    &&$0.842$&&  $3.023$   & & $0.844$  \\
$0.0$ && $3.069$   &&$0.849$&& $3.069$    &&$0.849$&&  $3.069$   & & $0.849$  \\
\hline
\end{tabular}
\caption[]{Behavior of $\bar{g}(\sigma^{+},\cos \theta_{12})$ and $\bar{g}(\lambda \sigma^{-}, \cos \theta_{12})$ for $\cos \theta_{12}=-1,0,1$ corresponding to HH, X and HT configurations respectively. All values are as in the previous table.
\label{tab:gave}
}
\end{center}
\end{table}

\clearpage
\centerline{\bf Figure Captions}

\clearpage
\clearpage
\clearpage
\clearpage
\clearpage
\clearpage
\clearpage
\clearpage
\clearpage
\end{document}